\documentclass{article}

\usepackage{microtype}
\usepackage{graphicx}
\usepackage{subfigure}
\usepackage{booktabs} 
\usepackage{makecell}
\usepackage{multirow}
\usepackage{dsfont}

\usepackage{hyperref}

\usepackage[accepted]{icml2024}

\usepackage{amsmath}
\usepackage{amssymb}
\usepackage{mathtools}
\usepackage{amsthm}

\usepackage[capitalize,noabbrev]{cleveref}

\theoremstyle{plain}

\theoremstyle{definition}

\theoremstyle{remark}

\usepackage[textsize=tiny]{todonotes}

\icmltitlerunning{IOI: Invisible One-Iteration Adversarial Attack on No-Reference Image- and Video-Quality Metrics}

\begin{document}

\twocolumn[
\icmltitle{IOI: Invisible One-Iteration Adversarial Attack on No-Reference \\ Image- and Video-Quality Metrics}



\icmlsetsymbol{equal}{*}

\begin{icmlauthorlist}
\icmlauthor{Ekaterina Shumitskaya}{comp,sch,yyy}
\icmlauthor{Anastasia Antsiferova}{sch,comp}
\icmlauthor{Dmitriy Vatolin}{comp,sch,yyy}
\end{icmlauthorlist}

\icmlaffiliation{yyy}{Lomonosov Moscow State University, Moscow, Russia}
\icmlaffiliation{comp}{ISP RAS Research Center for Trusted Artificial Intelligence, Moscow, Russia}
\icmlaffiliation{sch}{MSU Institute for Artificial Intelligence Moscow, Russia}

\icmlcorrespondingauthor{Ekaterina Shumitskaya}{ekaterina.shumitskaya@graphics.cs.msu.ru}

\icmlkeywords{Machine Learning, ICML}

\vskip 0.3in
]



\printAffiliationsAndNotice{} 

\begin{abstract}
No-reference image- and video-quality metrics are widely used in video processing benchmarks. The robustness of learning-based metrics under video attacks has not been widely studied. In addition to having success, attacks on metrics that can be employed in video processing benchmarks must be fast and imperceptible. This paper introduces an Invisible One-Iteration (IOI) adversarial attack on no-reference image and video quality metrics. The proposed method uses two modules to ensure high visual quality and temporal stability of adversarial videos and runs for one iteration, which makes it fast. We compared our method alongside eight prior approaches using image and video datasets via objective and subjective tests. Our method exhibited superior visual quality across various attacked metric architectures while maintaining comparable attack success and speed. We made the code available on GitHub: \url{https://github.com/katiashh/ioi-attack}.
\end{abstract}

\section{Introduction}
\label{sec:introduction}

No-reference (NR) image- and video-quality assessment poses a significant challenge in computer vision. In contrast to full-reference (FR) quality metrics, NR metrics do not estimate the similarity of a distorted image or video to the original one but evaluate its visual appeal. The rapid integration of deep-learning-based NR image-and video-quality assessment metrics \cite{ying2020patches, talebi2018nima, Su_2020_CVPR, golestaneh2021no} led to the importance of investigating their vulnerabilities to transformations of input. One of the most common types of input transformations is adversarial attacks. In the case of image- and video-quality metrics, adversarial attacks are modifications of input images or videos that change the predicted quality score without significant influence on perceptual quality. Several studies \cite{zhang2022perceptual, korhonen2022adversarial, sang4112969generation, shumitskaya2022universal, DBLP:conf/iclr/ShumitskayaAV23, meftah2023evaluating, yang2024exploring} unveiled vulnerabilities in NR image quality metrics when exposed to adversarial attacks. 

NR metrics are employed in various image- and video-processing benchmarks, such as super-resolution \cite{ImageSRonPIRM, Khrulkov-2021-CVPR, 10.1016/j.cviu.2016.12.009}, video generation \cite{video-generation-paperswithcode}, video compression \cite{46070}. For some tasks, NR metrics show even better performance than FR metrics; for example, video super-resolution \cite{sr-metrics-benchmark} or video compression by new encoding standards such as H.266\slash VVC \cite{NEURIPS2022-59ac9f01}. The developers of video-processing algorithms can integrate adversarial attacks on quality metrics into their methods to achieve higher positions in public benchmarks. Today, such cheating can be detected in benchmarks that publish subjective comparisons along with objective ones. For example, in MSU Codec Comparison 2021 \cite{msu-subjective-video-codecs-benchmark-2021}, the leaderboard by a learning-based metric VMAF, which was shown to be vulnerable to attacks \cite{10.1145/3508259.3508272, zvezdakova2019hacking}, differs from a subjective one. Algorithms in these benchmarks compete in both visual quality and speed, and a high speed is essential for some real-life applications like universal encoding for video compression (from one to ten frames per second). Thus, to cheat in video processing benchmarks, it is profitable for an attacker to inject an imperceptible perturbation into the video without significantly decreasing the method speed.

In the literature, most of the existing approaches evaluate NR metrics robustness in the image domain. However, to evaluate the robustness of NR metrics for videos, an adversary has to satisfy several essential conditions, making such a task more challenging:
\begin{enumerate}
\item Quantitative success of an attack. For NR metrics, the success of an attack is measured by the amount of the metric's score change. Both decreasing and increasing a metric's score can be considered an attack; however, making a score higher has more practical applications.
\item High speed of an attack. The attack must operate at high speed for practical viability. A slow attack holds limited practical significance, as its integration into video processing algorithms will greatly slow down the method and lower its position in benchmarks. 
\item Temporal consistency of a transformed video. The per-frame implementation of adversarial attacks designed for images in videos leads to noticeable flickering effects that look suspicious in subjective comparisons.
\end{enumerate} 


Our research aims to investigate the potential of injecting fast, invisible and temporally consistent adversarial attacks on NR quality metrics. This paper introduces the Invisible One-Iteration (IOI) adversarial attack for images and videos. To achieve high attack speed, our method yields perturbation by calculating the gradient of an attacked model using one access to the model. 
We further show that a one-iteration attack for each frame is more efficient than a many-iteration attack applied to only some frames. To keep the temporal stability of a perturbed video, we make our attack invisible to the human eye by using weighting and frequency modules. The proposed attack operates in a white-box scenario, which does not limit its applicability for benchmarks: usually, a comparison methodology is known, and quality metrics are published for reproducibility. The primary contributions of this work can be summarized as follows:

\begin{itemize}
\item We propose an Invisible One-Iteration (IOI) adversarial attack that increases NR image- and video-quality metrics scores. It produces perturbations that are imperceptible and temporally stable in videos. The attack is fast: it does not require convergence and works efficiently with one iteration.
\item We propose a methodology for comparing adversarial attacks at NR metrics. It is based on aligning attack speed and relative metrics increase after the attack, yielding to comparing only objective and subjective quality of adversarial videos.
\item We conducted comprehensive experiments using two datasets and three NR models. Four quality metrics were used to demonstrate that the proposed attack generates adversarial images and videos of superior quality compared to prior methods. 
\item We conducted a subjective study on the proposed method's perceptual quality and temporal stability. A crowd-sourced subjective comparison with 685 subjects showed that the proposed attack produces adversarial videos of better visual quality than previous methods.
\end{itemize}
Our code is publicly available at \url{https://github.com/katiashh/ioi-attack}.

\section{Related Work}

Adversarial attacks usually have constraints on perturbations and minimize the $ L_{\infty} $ norm between adversarial examples and their original inputs. Other $ L_p $ norms \cite{su2019one, szegedy2013intriguing} are less common due to a higher computational complexity. We further consider methods that use only $ L_{\infty} $ constraint to work faster on high-dimensional data like videos.
Among such kinds of attacks, Goodfellow et al. proposed the \textbf{FGSM} method \cite{goodfellow_explaining_2015} that generates adversarial examples by leveraging gradients from the targeted model. Recently, \textbf{UAP} \cite{shumitskaya2022universal} and \textbf{FACPA} \cite{DBLP:conf/iclr/ShumitskayaAV23} attacks on NR image quality metrics have been proposed. These methods also used $ L_\infty $ norm constraints at the training stage. 

Numerous studies showed that $ L_p $ norms are not suitable as a distance metric to evaluate perceptual image quality \cite{sharif2018suitability, fezza2019perceptual, wang2004image, johnson_perceptual_2016, isola_image--image_2017}. Perturbations in images generated under $ L_p $ norm constraints often result in noisy pixels within smooth areas of the original image, which is easily perceptible to the human eye. 
Several adversarial attacks prioritize the visual quality of the generated adversarial images and use more sophisticated restrictions on perturbations than the bare implementation of $ L_p $ constraints.

Zhang et al. \yrcite{chen_advjnd_2020} introduces a novel approach \textbf{AdvJND} by incorporating just noticeable difference (JND) \cite{yang2005just} coefficients into the $ L_\infty $ norm constraint during adversarial example generation. These coefficients account for the human eye's ability to perceive the threshold of changes in an image. The authors employed I-FGSM and FGSM algorithms as baselines. To enhance the visual fidelity of adversarial images, they amplified original perturbations obtained from the FGSM or I-FGSM methods by scaled JND coefficients. 

\textbf{SSAH} \cite{luo_frequency-driven_2022} adversarial attack targets two objectives: semantic similarity of images and low-frequency constraint. The first component is crafted with a focus on image classification tasks. The second component adds perturbations within high-frequency regions by minimizing the difference between low-frequency information of adversarial and original images. Usually, it requires many iterations to converge and produce good visual quality. 

\textbf{Korhonen et al.} \yrcite{korhonen_adversarial_2022} introduced an iterative attack method targeting NR quality metrics, employing a Sobel filter to hide distortions within textured regions. In each iteration, the model's gradient under attack concerning the input is multiplied by a spatial activity map derived from the original image via the Sobel filter. This map highlights areas with substantial texture, enhancing the visual fidelity of the resulting adversarial images.

\textbf{Zhang et al.} \yrcite{zhang2022perceptual} introduced an iterative adversarial image crafting approach leveraging various FR metrics like Chebyshev distance, SSIM, LPIPS, and DISTS to manage visual distortions. Their method involved the iterative minimization of a loss function consisting of two components: the attacked loss and a loss based on some differential FR image quality metric.

Karli et al. \yrcite{karli_improving_2021} introduced the \textbf{Normalized Variance Weighting (NVW)} method aimed at amplifying perturbations within high-variance regions of images. This technique can be utilized alongside gradient attacks like FGSM or I-FGSM. Additionally, the authors proposed the LPIPS-minimization method, aiming to enhance perceptual quality by minimizing the LPIPS distance between the original and adversarial images while ensuring the classifier remains deceived. However, this minimization method is exclusively effective for discrete tasks, such as classification or detection, while applying it to quality metrics will lead to eliminating relative gain.

Table \ref{tab:related-works} summarizes existing attacks on image- and video-quality metrics. The primary drawback of the prior methods is their requirement to run the attack via many iterations with small steps. This leads to low attack speed, particularly on high-resolution video data. They can be used efficiently to attack videos only when applied to each $k$-th frame, which, as we further show, reduces relative gain and temporal consistency. 
\begin{table}[htb]
\caption{Comparison of existing adversarial attacks on image- and video-quality metrics regarding visual quality regulation features and the requirement to converge for an attack to succeed.}
\label{tab:related-works}
\begin{center}
\begin{small}
\begin{tabular}{lccc}
\toprule
 & \multicolumn{2}{c}{Visual quality} & Speed \\
Method & Weights & \makecell{Freq. reg.} & \makecell{Needn't \\ converg.} \\
\midrule
FGSM \yrcite{goodfellow_explaining_2015} & $\times$ & $\times$ & $\surd$ \\
UAP \yrcite{shumitskaya2022universal} & $\times$ & $\times$ & $\surd$ \\
FACPA \yrcite{DBLP:conf/iclr/ShumitskayaAV23} & $\times$ & $\times$ & $\surd$ \\
AdvJND \yrcite{chen_advjnd_2020} & JND map & $\times$ & $\surd$ \\
SSAH \yrcite{luo_frequency-driven_2022} & $\times$ & DWT  & $\times$\\
\makecell[l]{Korhonen et al. \\ \yrcite{korhonen_adversarial_2022}} & Sobel map & $\times$ & $\surd$ \\
\makecell[l]{Zhang et al. \\ \yrcite{zhang2022perceptual}} & $\times$ & $\times$ & $\times$\\
NVW \yrcite{karli_improving_2021} & Local STD & $\times$ & $\surd$ \\
\midrule
IOI (proposed) & \makecell{Local \\ STD-based} & FFT & $\surd$ \\
\bottomrule
\end{tabular}
\end{small}
\end{center}
\end{table}


\section{Proposed method}

\subsection{Problem formulation}

The adversarial attack on the NR quality metric $M$ is usually modelled as follows:
\begin{equation}
\underset{I^a}{\mathrm{argmax}} \{M(I^a) - M(I) \}, \lVert I^a - I \rVert _{p} \leq \epsilon,
\end{equation}
where $ I $ is a clear video frame, $ I^a $ is an attacked frame, $ \epsilon $ is a small constant. 

To ensure a high visual quality of the perturbed image, instead of using  $l_p$ norms that are inefficient for this task, we formulate the problem as follows:

\begin{equation}
\begin{array}{cc}
\underset{I^a}{\mathrm{argmax}} \{M(I^a) - M(I) \}, \\
\frac{L_f(I^a) \cdot L_f(I)}{||L_f(I^a)||_2 ||L_f(I)||_2} \ge 1 - \epsilon^*,
\end{array}
\end{equation}
where $L_f$ is a low-frequency filter, $ \epsilon^* $ is a small constant. When low-frequency components of the clear and adversarial images are closely aligned, the perturbations mainly affect high-frequency areas. As a result, distortions in adversarial images are nearly invisible to the human eye, according to the contrast masking theory of human vision \cite{legge1980contrast}. For measuring the difference between attacked and original images/frames during a perturbation construction, we use adversarial mean absolute error in the frequency domain ($MAE^*$), which is calculated as follows:
\begin{equation}
MAE^*(I^a, I) = \frac{1}{HW} \displaystyle\sum_{i=0}^{(H-1)} \sum_{j=0}^{(W-1)} |I^{a*}_{ij} - I_{ij}^*|,
\label{eq:mae}
\end{equation}
where $I^{a*}_{ij}$ and $ I_{ij}^* $ are Fast Fourier Transform (FFT) coefficients of attacked and original images correspondingly, $H$ and $W$ -- image dimensions.

As discussed in the introduction, we focus on the scenario where metric scores increase after an attack. In some other studies, NR metrics are modified in two directions, and the attack success is measured by a decrease in the metric's correlation with subjective quality \cite{zhang_perceptual_2022, zhang2023vulnerabilities}. While this approach holds theoretical significance in evaluating the stability of metric scores, it carries certain limitations. Only increasing the target metric score will keep the same correlation with subjective quality; thus, an attack remains undetected. Instead of using correlation as a measure of NR metrics' adversarial robustness, we evaluate attack success using relative gain (RG):
\begin{equation}
RG = \frac{M(I^a) - M(I)}{M_{range}}
\label{eq-rg}
\end{equation}
where $M_{range}$ is the range of scores produced by $M$.
 
\subsection{One-iteration attack}

\begin{figure*}[ht]
\begin{center}
\centerline{\includegraphics[width=\linewidth]{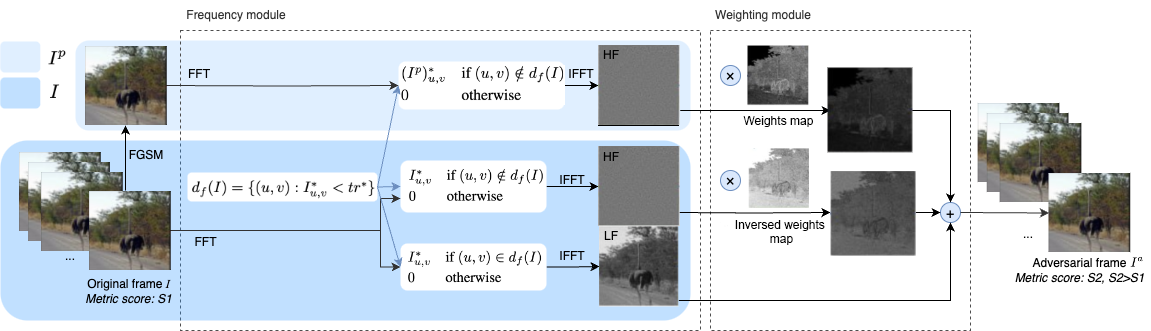}}
\caption{An overview of the proposed IOI adversarial attack. $I$ is stands for input image, $I^p$ -- FGSM attacked image and $I^a$ -- the final IOI attacked image. Weights map is calculated using formula \ref{eq:weights1}.}
\label{fig:main_sch}
\end{center}
\end{figure*}

Figure \ref{fig:main_sch} provides an overview of our method. Initially, the proposed attack perturbs the image using a baseline gradient attack. Subsequently, it processes the perturbed image using two modules to enhance the visual quality of an adversarial image/video: the frequency module and the weighting module. For the baseline gradient attack, we employ FGSM \cite{goodfellow_explaining_2015}: 
\begin{equation}
I^p = I + \epsilon * sign(\nabla_{I} M(I)) \\
\label{eq:fgsm}
\end{equation}

\subsection{Frequency module}

The frequency module extracts features from the original $I$ and perturbed $I^p$ images for further processing. It decomposes them into low-frequency (LF) and high-frequency (HF) components using the Fast Fourier Transform (FFT) and Inverse Fast Fourier Transform (IFFT). 
The LF component keeps fundamental content information, while the HF component contains noise and texture details. A threshold $tr^*$ for dividing frequencies in f\% highest and (1-f)\% the lowest coefficients is used to calculate indexes $d_f(I)$ of the highest FFT coefficient of the original image $I$ (Equation \ref{eq:freq2}). 
\begin{equation}
\begin{array}{cc}
tr^* = \underset{tr}{\mathrm{argmin}} |f - \frac{\sum_{u=0}^{H-1}\sum_{v=0}^{W-1} \mathds{1} [I^*_{u,v} > tr]}{HW}| \\
d_f(I) = \{ (u, v) : I^*_{u,v} > tr^* \}
\label{eq:freq2}
\end{array}
\end{equation}
We use $d_f(I)$ to extract LF and HF components from the original image, as shown in Equation \ref{eq:freq32}. These components are further utilized in the weighting module to enhance the visual quality of adversarial video frames. HF components of the perturbed image $I^p$ are extracted using indexes $d_f(I)$ from the original image. Thus, the perturbations will be transferred into HF areas of the original image (Equation \ref{eq:freq31}). The parameter $f$ allows fine-tuning the visibility of perturbations and relative attack gain at different levels. Lower values of $f$ result in more visible perturbations within the frequency module and higher relative gain. 
\begin{equation}
\begin{array}{cc}
L^{d_f(I)}_{u,v}(I) =
  \begin{cases}
    I^*_{u,v}       &  \text{if } (u,v) \in d_f(I)\\
    0  &  \text{otherwise } \\
  \end{cases} \\
H^{d_f(I)}_{u,v}(I) =
  \begin{cases}
    I^*_{u,v}       &  \text{if } (u,v) \notin d_f(I)\\
    0  &  \text{otherwise } \\
  \end{cases} \\
\label{eq:freq32}
\end{array}
\end{equation}
\begin{equation}
\begin{array}{cc}
 H^{d_f(I)}_{u,v}(I^p) =
  \begin{cases}
    (I^p)^*_{u,v}       &  \text{if } (u,v) \notin d_f(I)\\
    0  &  \text{otherwise } \\
  \end{cases} \\
\label{eq:freq31}
\end{array}
\end{equation}

\subsection{Weighting module}

\begin{figure}[!h]
\begin{center}
\centerline{\includegraphics[width=1.7in]{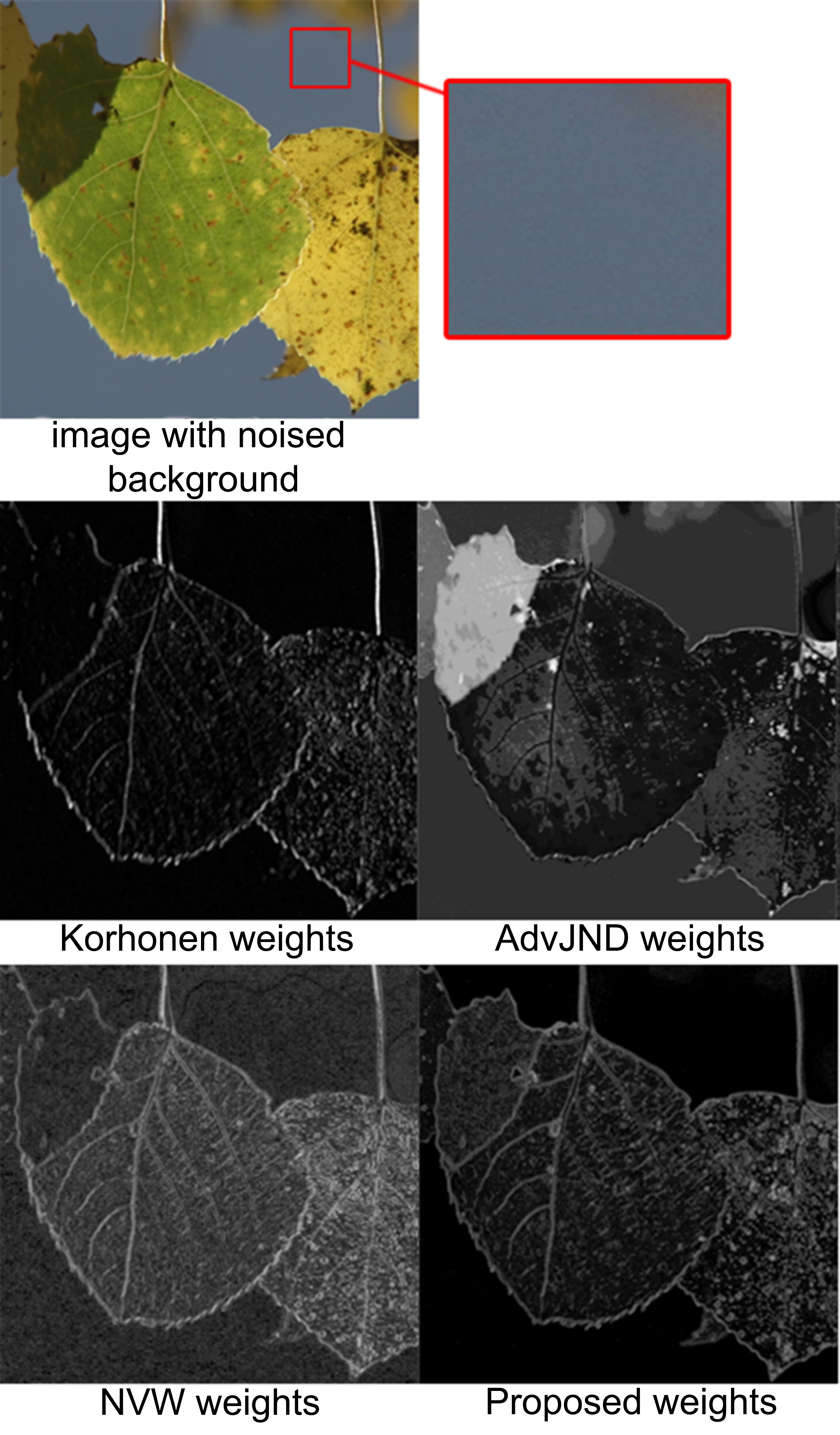}}
\caption{Comparison of weights used in prior and proposed methods. NVW and AdvJND assign non-zero weights for a background. Korhonen et al.'s weight total area is relatively small.}
\label{fig:weights}
\end{center}
\end{figure}

Various image processing applications \cite{lin2005visual, liu2010just} operate under the assumption that distortions in low-variance regions are more visible to the human eye than in high-variance areas. The weighting module generates a weighting map for the input image based on the variance map. A similar approach was used in previous studies \cite{croce2019sparse, karli_improving_2021} for generating adversarial examples. We introduce additional features to enhance the weights obtained from the variance map method. The proposed weights map generation method is described in Equation \ref{eq:weights1}. Initially, we compute local variance and local mean maps for an image $ x $, determining standard deviations and means for both axes using a window size of 3 for each colour channel. Next, we derive the relative local variance of the image by dividing the local variance map by the local mean map and normalizing the weights to the range [0, 1]. In the next step, we zero out 1\% and compute the square root of the weights.
\begin{equation}
\label{eq:weights1}
\begin{array}{cc}
\sigma_{i, j} = \sqrt{ \frac{\sum x_{i,j}^2 }{n} - (\frac{\sum x_{i,j}}{n})^2 }, m_{i, j} = \frac{\sum x_{i,j}}{n} \\
\gamma = \frac{\sigma}{m}, \text{ } \gamma_{max} = \underset{i,j}\max{(\gamma_{i,j})}, \text{ } \gamma_{norm}=\frac{\gamma}{\gamma_{max}}\\
w_{i,j} = 
 \begin{cases}
   \sqrt{(\gamma_{norm})_{i,j}} \text{ , if } (\gamma_{norm})_{i,j} \ge 0.01 \\
   0 \text{ , if } (\gamma_{norm})_{i,j} < 0.01
 \end{cases} \\
\end{array}
\end{equation}
Figure \ref{fig:weights} illustrates the weights map derived by the proposed method and three prior methods (Korhonen et al. \yrcite{korhonen2022adversarial}, AdvJND \cite{chen_advjnd_2020}, and NVW \cite{karli_improving_2021}). NVW and AdvJND methods assign non-zero weights for a noisy background, often resulting in visible distortions within smooth regions. The image area covered by non-zero weights in Korhonen et al.'s map is small, potentially limiting the strength of the attack it can generate. 

We use the proposed weights map to guide the weighting of adversarial image HF components. As a result, the ultimate perturbation remains absent in smooth areas of an image. The construction of the final adversarial image involves the composition of three elements: the original LF component, the perturbed HF component multiplied by the weights map, and the original HF component multiplied by inverse weights:
\begin{equation}
I^a = L^{d_f(I)}_c(I) + w H^{d_f(I)}_c(I^p) + (1 - w) H^{d_f(I)}_c(I)
\label{eq:ia-formula}
\end{equation}

\subsection{Mathematical properties}

This section provides theoretical restrictions of the generated adversarial images or video frames by the proposed method. 

\textbf{Theorem 1.} \textit{Let $I \text{ and } I^p$ be original and perturbed image correspondingly, $I^a$ -- adversarial image after IOI attack that is based on $I^p$ with truncating parameter $f$. Then inequality \ref{eq:formula} is correct, where $MAE^*(\cdot, \cdot)$ is given by Equation \ref{eq:mae}.}
\begin{equation}
\begin{array}{l}
||I^a - I||_{\infty} \leq (1-f) MAE^*(I^p, I)
\end{array}
\label{eq:formula}
\end{equation}
\textbf{Proof.} Each element of the $(I^a - I)$ is estimated using Equation \ref{eq:ia-formula} and representation of high-frequency FFT components as re-transformed 2D discrete Fourier transform without f\% of highest frequencies. The complete proof is presented in the Appendix \ref{appendix-proof}.

The statement above demonstrates that the proposed method guarantees theoretical restriction of $l_\infty$ norm of the adversarial image, which depends on initial attack strength expressed in $MAE^*$ and parameter $f$ for truncating frequencies. Higher $f$ leads to saving more frequencies and providing better visual quality of the generated IOI adversarial image/video.

\section{Experiments}
We compared our method to the previous approaches targeting three learning-based NR image- and video-quality metrics on two datasets of images and videos. 

\textbf{Datasets.}
NIPS2017 image dataset \yrcite{nips-dataset} was used to evaluate attacks on three NR metrics. It includes 1,000 images of a 299$\times$299 resolution. For evaluating methods on videos, we used 12 videos with 1280$\times$720 resolution from the DERF dataset \yrcite{derf-dataset}. Descriptions of these videos are available in the Appendix \ref{sec:appendix-video-desc}. We extracted 75 frames from each original video and saved an attacked video with a frame rate of 25 frames per second, resulting in 3-second videos. The datasets licenses allow usage for research purposes.

\textbf{Attacked models.} 
For experiments on images, we selected PaQ-2-PiQ \cite{ying2020patches}, Hyper-IQA \cite{Su_2020_CVPR}, and TReS \cite{golestaneh2021no} NR models. For videos, we attacked the PaQ-2-PiQ \cite{ying2020patches} metric. These metrics were chosen to cover different architectures. PaQ-2-PiQ \cite{ying2020patches} employs RoIPool layers, which allows the flexibility to aggregate at different scales. Hyper-IQA \cite{Su_2020_CVPR} utilizes ResNet50 for semantic feature extraction with further processing in the proposed Content Understanding Hyper Network. TReS \cite{golestaneh2021no} is based on transformer architecture.


\textbf{Methodology.} 
To evaluate the efficiency of adversarial attacks, we considered three factors: relative gain (Equation \ref{eq-rg}), speed and the objective or subjective quality of perturbed images/videos. 
To compare these three factors, we fixed two of them by aligning the speed and relative gain of all methods and compared the objective and subjective visual quality of adversarial images/videos. Since most of the attack time is spent on backpropagation, we executed each attack for one iteration to standardize the speed of all attacks. We also conducted additional experiments with multiple iterations, results presented in the Appendix \ref{appendix-mult-iters}. In a one-iteration comparison, we combined the results of Zhang et al. and SSAH attacks with FGSM attacks in all tables because one iteration of these attacks is equivalent to one iteration of FGSM. Zhang et al. use an FR quality metric to preserve image fidelity, and SSAH uses the distance between low-frequency information of two images. Since the distorted image used in both methods appears only during the first iteration, these fidelity-preserving components yield zero gradients. 


We employed an automatic process described in Algorithm \ref{alg:example} to ensure equal relative gain. Each attack has a parameter to regulate its strength. 
We denote this parameter as $lr$. Initially, we ran the proposed method with fixed parameters ($lr=0.1$ and $f=0.07$ for image data and $lr=0.1$ and $f=0.05$ for video data). Then, for each other attack and each image/video, we searched the minimal $lr$ parameter to achieve the same relative gain. 
The search process also halted if the reached relative gain did not improve for $n=5$ search iterations. 
For videos, the quality score was calculated as the mean of quality scores on each frame (PaQ-2-PiQ \cite{ying2020patches}, Hyper-IQA \cite{Su_2020_CVPR}, and TReS \cite{golestaneh2021no} are metrics that run per-frame on videos).

\begin{algorithm}[htb]
   \caption{Relative gain aligning}
   \label{alg:example}
\begin{algorithmic}
\STATE  $ \textbf{Inputs: }  \text{data element } \mathbf{X} \text{, target relative gain } RG_{t},$
$\text{adversarial attack } \mathbf{Adv}, \text{ attacked model } \mathbf{M}, $
$ \text{search step } d, \text{ stop parameter } n, \text{ range of } \mathbf{M} \text{ } M_{range} $
\STATE $ \textbf{Output: }  \text{attacked data element }  \mathbf{X_{adv}} $

\STATE $ lr = 0\text{, } counter = 0\text{, } RG_{prev} = 0\text{, } flag = False  $

\STATE $ \textbf{while not } flag \textbf{ do}$

\STATE \hspace{0.5cm} $ \mathbf{X_{adv}} = \mathbf{Adv}(\mathbf{X}, \mathbf{M}, lr)$

\STATE \hspace{0.5cm} $ RG = \frac{\mathbf{M(X_{adv})} - \mathbf{M(X)}}{M_{range}}$

\STATE \hspace{0.5cm} $ \textbf{if } RG \ge RG_{t} \textbf{ do } flag = True$


\STATE \hspace{0.5cm} $ \textbf{if } RG \le RG_{prev} \textbf{ do } counter = counter + 1$


\STATE \hspace{0.5cm} $ \textbf{if } counter == n \textbf{ do } flag = True$


\STATE \hspace{0.5cm} $ lr = lr + d, \text{ } RG_{prev} = RG$


\STATE $ \textbf{end while}$
\end{algorithmic}
\end{algorithm}

\textbf{Quality metrics and subjective study.}
We compared the objective quality of adversarial images and videos using four FR image- and video-quality metrics: PSNR, SSIM \cite{wang2004image}, VIF \cite{sheikh2006image}, and LPIPS \cite{zhang2018perceptual}.

We conducted a crowd-sourced subjective comparison using Subjectify.us \cite{subjectify} to get subjective scores for adversarial videos. Original and adversarial videos were compressed using the x264 video codec with a CRF value 16 (preset ``Medium''). Each participant was asked to choose the video of the superior visual quality from a random pair of videos shown sequentially. An option ``Can't choose'' was also available for them. Videos were pre-downloaded in the browser to prevent delays in playback, and participants had the flexibility to replay the videos multiple times. Each participant compared 12 video pairs; two of the 12 pairs were special verification pairs with apparent differences in visual quality. Answers from 200 participants who failed the verification were excluded. We collected 8220 responses from 685 participants who passed verification and calculated subjective scores using the Bradley-Terry model \cite{bradley1952rank}. More details about the subjective experiment setup are presented in the Appendix \ref{appendix-subj-comp}.

\begin{table*}[h]
\caption{The objective quality of adversarial images generated by existing and proposed methods on the NIPS2017 image dataset \yrcite{nips-dataset} for three attacked models: PaQ-2-PiQ \cite{ying2020patches}, Hyper-IQA \cite{Su_2020_CVPR}, and TReS \cite{golestaneh2021no}. The table presents FR metrics scores for adversarial images averaged across the dataset, with aligned relative gain and 95\% confidence intervals. Each attack run for one iteration.}
\label{tab:res}
\begin{center}
\begin{small}
\begin{tabular}{llrrrrr}
\toprule
Attacked model & Method & SSIM $\uparrow$ & PSNR $\uparrow$ & VIF $\uparrow$ & LPIPS $\downarrow$  \\
\midrule
\multirow{7}{*}{\makecell{PaQ-2-PiQ \\\yrcite{ying2020patches}}}& \makecell[l]{FGSM \yrcite{goodfellow_explaining_2015}, SSAH \yrcite{luo_frequency-driven_2022}, \\ Zhang et al. \yrcite{zhang_perceptual_2022}} & 0.884$\pm$0.007 & \underline{33.6$\pm$0.3} & 0.635$\pm$0.010 & 0.134$\pm$0.009  \\
& NVW \yrcite{karli_improving_2021} & \underline{0.897$\pm$0.007} & \textbf{34.7$\pm$0.5} & \underline{0.648$\pm$0.011} & \underline{0.120$\pm$0.008}  \\
& Korhonen et al. \yrcite{korhonen_adversarial_2022} & 0.872$\pm$0.008 & 33.1$\pm$0.3 & 0.617$\pm$0.011 & 0.151$\pm$0.011 \\
& AdvJND \yrcite{chen_advjnd_2020} & 0.740$\pm$0.008 & 29.5$\pm$0.2 & 0.384$\pm$0.008 & 0.208$\pm$0.007 \\
& UAP \yrcite{shumitskaya2022universal} & 0.737$\pm$0.004 & 26.3$\pm$0.2 & 0.371$\pm$0.004 & 0.314$\pm$0.005 \\
& FACPA \yrcite{DBLP:conf/iclr/ShumitskayaAV23} & 0.863$\pm$0.003 & 30.5$\pm$0.2 & 0.539$\pm$0.005 & 0.182$\pm$0.004 \\
& IOI (ours) & \textbf{0.950$\pm$0.002} & 33.4$\pm$0.2 & \textbf{0.695$\pm$0.005} & \textbf{0.059$\pm$0.003} \\
\midrule
\multirow{7}{*}{\makecell{Hyper-IQA \\ \yrcite{Su_2020_CVPR}}}& \makecell[l]{FGSM \yrcite{goodfellow_explaining_2015}, SSAH \yrcite{luo_frequency-driven_2022}, \\ Zhang et al. \yrcite{zhang_perceptual_2022}} & 0.746$\pm$0.017 & 30.6$\pm$0.6 & 0.542$\pm$0.019 & 0.326$\pm$0.023 \\
& NVW \yrcite{karli_improving_2021} & 0.801$\pm$0.015 & 33.4$\pm$0.7 & 0.610$\pm$0.019 & 0.255$\pm$0.021 \\
& Korhonen et al. \yrcite{korhonen_adversarial_2022} & 0.765$\pm$0.016 & 31.1$\pm$0.6 & 0.562$\pm$0.019 & 0.303$\pm$0.022 \\
& AdvJND \yrcite{chen_advjnd_2020} & \underline{0.909$\pm$0.004} & \textbf{37.1$\pm$0.3} & \underline{0.660$\pm$0.011} & \underline{0.073$\pm$0.005}  \\
& UAP \yrcite{shumitskaya2022universal} & 0.545$\pm$0.010 & 21.4$\pm$0.3 & 0.192$\pm$0.007 & 0.447$\pm$0.008 \\
& FACPA \yrcite{DBLP:conf/iclr/ShumitskayaAV23} & 0.627$\pm$0.008 & 24.8$\pm$0.2 & 0.270$\pm$0.007 & 0.299$\pm$0.007 \\
& IOI (ours) & \textbf{0.952$\pm$0.002} & \underline{33.5$\pm$0.2} & \textbf{0.722$\pm$0.005} & \textbf{0.058$\pm$0.003}  \\
\midrule
\multirow{7}{*}{\makecell{TReS \\ \yrcite{golestaneh2021no}}} & \makecell[l]{FGSM \yrcite{goodfellow_explaining_2015}, SSAH \yrcite{luo_frequency-driven_2022}, \\ Zhang et al. \yrcite{zhang_perceptual_2022}} & 0.876$\pm$0.011 & 35.9$\pm$0.4 & 0.719$\pm$0.015 & 0.134$\pm$0.013 \\
& NVW \yrcite{karli_improving_2021} & 0.902$\pm$0.010 & \underline{37.7$\pm$0.5} & \underline{0.754$\pm$0.014} & 0.107$\pm$0.011  \\
& Korhonen et al. \yrcite{korhonen_adversarial_2022} & 0.888$\pm$0.011 & 36.3$\pm$0.4 & 0.734$\pm$0.015 & 0.123$\pm$0.013 \\
& AdvJND \yrcite{chen_advjnd_2020} & \underline{0.915$\pm$0.006} & \textbf{39.1$\pm$0.4} & 0.736$\pm$0.013 & \underline{0.064$\pm$0.006}  \\
& UAP \yrcite{shumitskaya2022universal} & 0.445$\pm$0.008 & 17.5$\pm$0.1 & 0.120$\pm$0.003 & 0.715$\pm$0.008 \\
& FACPA \yrcite{DBLP:conf/iclr/ShumitskayaAV23} & 0.611$\pm$0.007 & 23.4$\pm$0.2 & 0.221$\pm$0.007 & 0.530$\pm$0.011 \\
& IOI (ours) & \textbf{0.945$\pm$0.002} & 33.4$\pm$0.2 & \textbf{0.756$\pm$0.005} & \textbf{0.059$\pm$0.003} \\
\bottomrule
\end{tabular}
\end{small}
\end{center}
\end{table*}

\section{Results}

Table \ref{tab:res} compares the proposed IOI adversarial attack and eight prior attacks at one iteration on the NIPS2017 image dataset \yrcite{nips-dataset}. The comparison involves three attacked image quality models, and the relative gain is aligned using the proposed Algorithm \ref{alg:example}. On average, the relative gain achieved by all attacks was 7.7\% for all models.

The proposed IOI method showed higher SSIM, VIF, and LPIPS scores for all attacked NR metrics. The PSNR score of our attack method is lower than that of other methods, which means that IOI changes more information in images; however, the perturbations are hidden in the images' texture/contrast regions. AdvJND method showed promising results for attacking Hyper-IQA and TReS models but failed on PaQ-2-PiQ. NVW performed well on PaQ-2-PiQ and TReS. 

Table \ref{tab:res_vid} contains the results of objective comparison on 12 videos from the DERF dataset \yrcite{derf-dataset} and attacking the PaQ-2-PiQ \cite{ying2020patches} metric. We applied attacks with greater intensity on videos to enhance distinguishability for further subjective comparison, so the average relative gain was 15.3\%. The objective results are similar to the comparison on the images: IOI outperformed prior methods on SSIM, VIF, and LPIPS metrics and showed comparable PSNR scores. The per-video results are in the Appendix \ref{sec:appendix-per-video}.

\begin{figure*}[ht]
\begin{center}
\centerline{\includegraphics[width=5.4in]{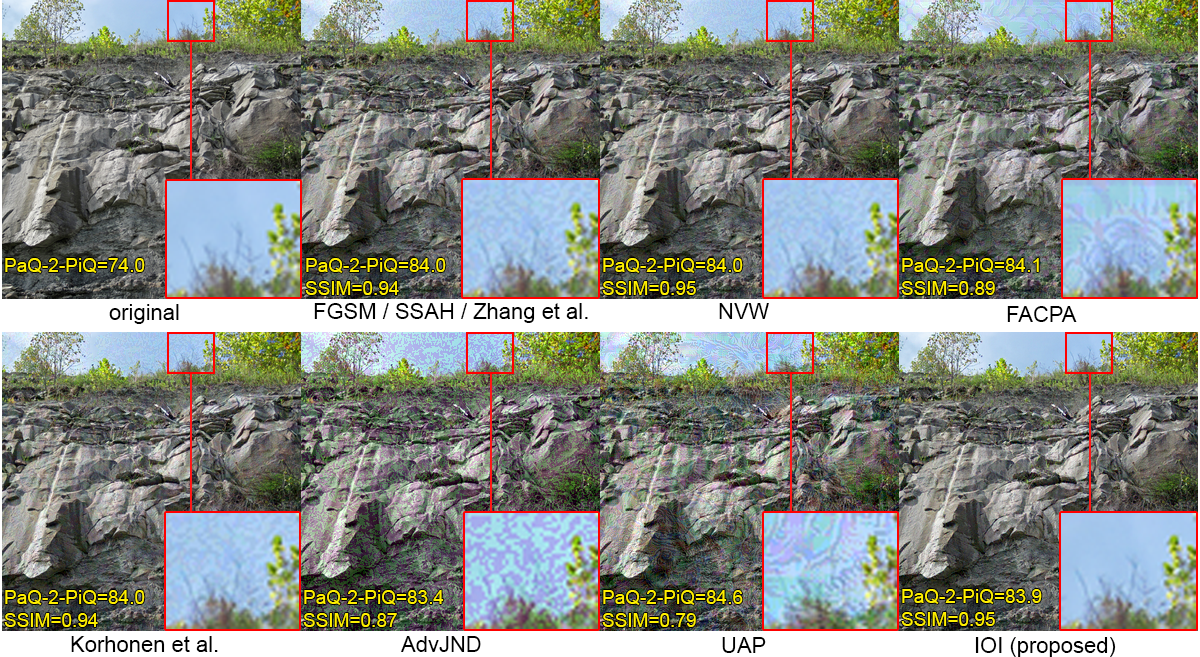}}
\caption{Comparison of adversarial images generated using FGSM \yrcite{goodfellow_explaining_2015}, SSAH \yrcite{luo_frequency-driven_2022}, Zhang et al. \yrcite{zhang_perceptual_2022}, NVW \yrcite{karli_improving_2021}, Korhonen et al. \yrcite{korhonen_adversarial_2022}, AdvJND \yrcite{chen_advjnd_2020}, UAP \yrcite{shumitskaya2022universal}, FACPA \yrcite{DBLP:conf/iclr/ShumitskayaAV23} and IOI (ours) attack methods when attacking PaQ-2-PiQ \yrcite{ying2020patches} NR quality metric at one iteration with relative gain aligned by Algorithm \ref{alg:example}.}
\label{fig:ex1}
\end{center}
\end{figure*}

\textbf{Subjective comparison}.
The subjective scores obtained from pairwise comparisons showed that the IOI attack generates adversarial videos of better visual quality: it holds a quality of 2.97, while other methods' scores are below 2.16. Confidence intervals for IOI and other methods do not intersect. The intervals intersect for some prior methods, since their adversarial videos are noisy and flickering, making their subjective quality difficult to rank.
The video with the highest distinguishability was ``Blue Sky'' with tree branches swaying against the smooth sky. The video with the lowest distinguishability was ``Rush Hour'' with a highly noisy background. As shown in Figure \ref{fig:ex1}, FGSM, NVW, and Korhonen et al. methods generate imperceptible distortion on the stone region but fail to suppress the perturbation on the sky background. AdvJND, UAP and FACPA cause visible distortions on the whole image. To produce the same relative gain as IOI using one iteration, $lr$ for other methods, was high, yielding visible perturbations. As shown in the Appendix \ref{appendix-mult-iters}, all methods (except FGSM, UAP and FACPA) produce almost equivalent results at 20 iterations. But at one iteration, there is a crucial difference.
The videos used for the comparison are available at \url{https://github.com/katiashh/ioi-attack}.

\begin{table*}[tb]
\caption{Subjective comparison results on 12 videos from the DERF dataset \yrcite{derf-dataset}. Adversarial videos generated for PaQ-2-PiQ model \cite{ying2020patches} at equal speed and relative gain of all attacks. The table presents averaged quality metrics and subjective scores with 95\% confidence intervals. Each attack runs for one iteration on each video frame.}
\label{tab:res_vid}
\begin{center}
\begin{small}
\begin{tabular}{llrrrrr}
\toprule
Method & SSIM $\uparrow$ & PSNR $\uparrow$ & VIF $\uparrow$ & LPIPS $\downarrow$ & \makecell{Subjective \\ score $\uparrow$} \\
\midrule
\makecell{FGSM \yrcite{goodfellow_explaining_2015}, SSAH \yrcite{luo_frequency-driven_2022}, \\ Zhang et al. \yrcite{zhang_perceptual_2022}} & 0.859$\pm$0.005 & 33.1$\pm$0.2 & 0.555$\pm$0.007 & 0.195$\pm$0.006 & 1.95$\pm$0.16 \\
NVW \yrcite{karli_improving_2021} & 0.871$\pm$0.005 & 33.4$\pm$0.2 & 0.570$\pm$0.007 & 0.178$\pm$0.006 & \underline{2.16$\pm$0.16} \\
Korhonen et al. \yrcite{korhonen_adversarial_2022}  & 0.855$\pm$0.005 & 33.0$\pm$0.2 & 0.550$\pm$0.007 & 0.204$\pm$0.007 & 2.06$\pm$0.16 \\
AdvJND \yrcite{chen_advjnd_2020}  & 0.848$\pm$0.005 & \textbf{34.5$\pm$0.2} & 0.516$\pm$0.008 & \underline{0.153$\pm$0.006} & 1.76$\pm$0.16 \\
UAP \yrcite{shumitskaya2022universal} & 0.809$\pm$0.003 & 29.8$\pm$0.2 & 0.450$\pm$0.003 & 0.301$\pm$0.004 & 0.19$\pm$0.19 \\
FACPA \yrcite{DBLP:conf/iclr/ShumitskayaAV23} & \underline{0.887$\pm$0.002} & 32.9$\pm$0.2 & \underline{0.578$\pm$0.004} & 0.207$\pm$0.003 & 0.87$\pm$0.17 \\
IOI (ours) & \textbf{0.941$\pm$0.016} & \underline{34.3$\pm$1.7} & \textbf{0.669$\pm$0.046} & \textbf{0.098$\pm$0.030} & \textbf{2.97$\pm$0.16} \\
\bottomrule
\end{tabular}
\end{small}
\end{center}
\end{table*}

\section{Discussion}

\textbf{Different frame frequency and attack success.}
We conducted additional experiments to show the importance of a one-iteration setup when attacking NR quality metrics for videos. This section demonstrates that employing a single iteration for each frame produces superior results compared to the sporadic application of multiple iterations, such as ten iterations for every tenth frame. We selected the PaQ-2-PiQ \yrcite{ying2020patches} NR metric, ``Controlled Burn'' video from the DERF dataset \yrcite{derf-dataset}, and applied the I-FGSM attack \cite{kurakin2018adversarial}. I-FGSM is an extension of FGSM, involving multiple iterations.

Let $n$ represent the number of iterations, $\epsilon$ a small constant, $I$ the original video frame, and $M$ the target NR quality metric. The $k$-th iteration of I-FGSM is formulated as shown in Equation \ref{eq:ifgsm} ($k \in [0, n]$, $I^p_0 = I$).

\begin{equation}
\begin{array}{cc}
I^p_{k+1} = I^p_k + \frac{\epsilon}{n} * sign(\nabla_{I^p_k} M(I^p_k)) \\
\end{array}
\label{eq:ifgsm}
\end{equation}

\begin{figure*}[ht]
\begin{center}
\centerline{\includegraphics[width=\linewidth]{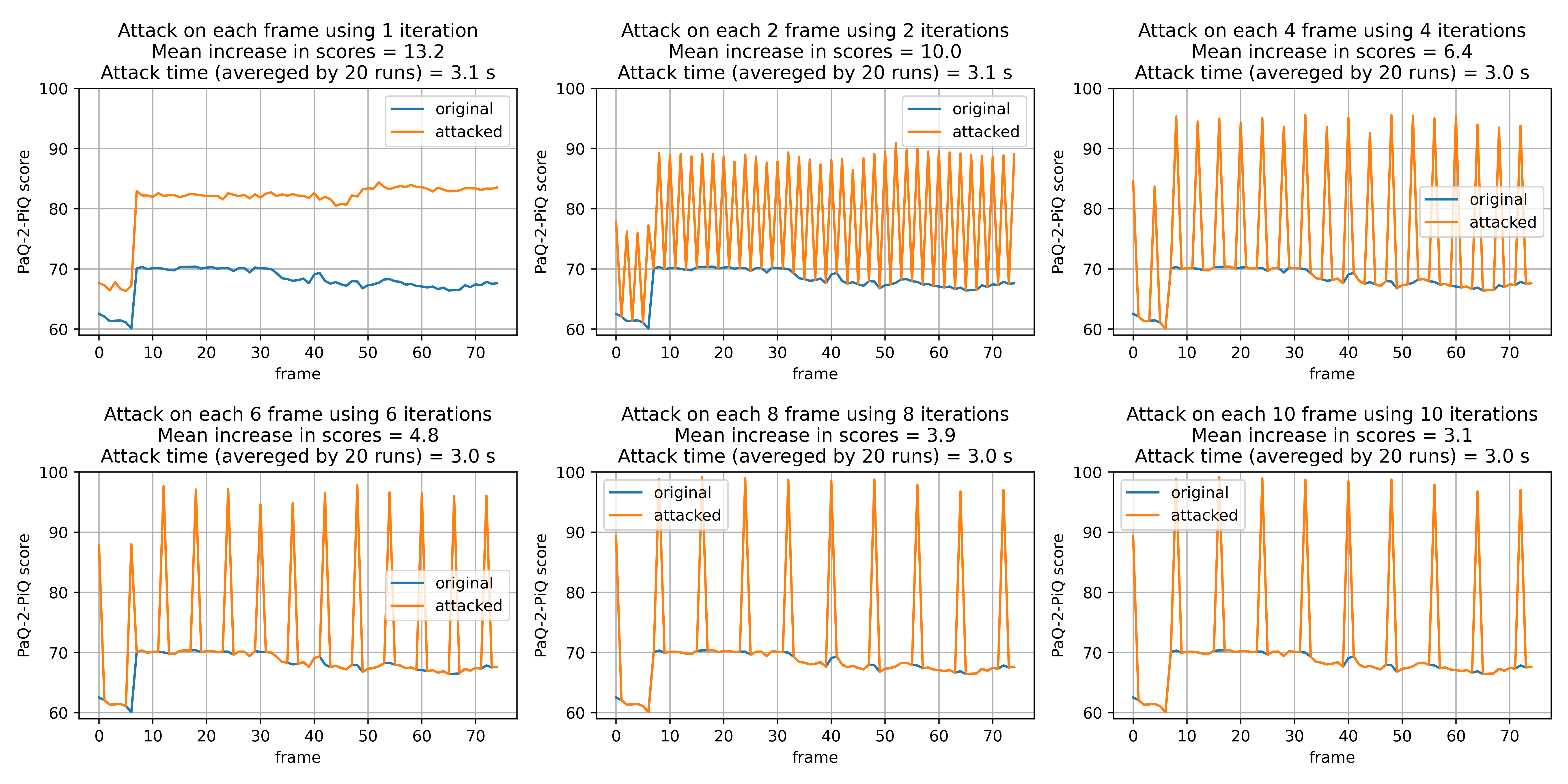}}
\caption{Results of experiments when attacking PaQ-2-PiQ \yrcite{ying2020patches} NR metric on the ``Controlled Burn'' video through I-FGSM attack \yrcite{kurakin2018adversarial} with different numbers of iterations and altered frames.}
\label{fig:c_burn_diff_runs}
\end{center}
\end{figure*}

We executed I-FGSM for different iteration counts: $n=1$, $n=2$, $n=4$, $n=6$, $n=8$, and $n=10$. Only some frames underwent attack in each experiment, specifically $\frac{1}{n}$. The selection of frames for the attack was done uniformly. The results of these experiments are illustrated in Figure \ref{fig:c_burn_diff_runs}. As the number of iterations in the attack increases, there is a corresponding rise in the attack's relative gain on each particular frame. However, evaluating the overall relative gain involves averaging relative gains across all frames. The optimal averaged relative gain occurs when each frame is attacked with just one iteration, and this gain decreases monotonically with the increase in the parameter $n$. We also measured the computation time for attacks with different values of $n$, which remained nearly constant at 3 seconds for all attacks. This consistency arises from the fact that each experiment's total number of adversarial iterations was the same.

Compared with other values of $n$, we can see that a one-iteration attack yields superior averaged relative gain within the same attack time. This highlights that attack on video quality metrics differ from the classification task, where an attacker can affect only several frames to fool the classifier. From the results of this experiment, we can conclude that the effectiveness of adversarial attacks for video quality metrics is defined by their effectiveness at a one-iteration setup. 

\textbf{Speed of the proposed method.} The PyTorch realization of the IOI attack allows reaching 8 fps on the NVIDIA Tesla T4 GPU. Details presented in Appendix \ref{app:one-it-speed}.


\textbf{IOI performance under defences}.
We did additional experiments (Table \ref{tab:disc_compress}) to check the robustness of the proposed method to three adversarial defences: video compression \cite{shaham2018defending}, random crop and resize used in \cite{shumitskaya2023towards} for NR metrics. Defences were evaluated for videos from the DERF dataset. Although the proposed method affects only high-frequency information, video compression reduced relative gain only by 2.4\%. Random cropping confuses the attack and reduces relative gain approximately two times. Frames resizing almost completely mitigates the relative gain from 14.6\% to 1\%; however, an NR metric increase by 1\% is still significant for benchmarks; sometimes, teams compete to achieve a 0.1\% metric increase to win the competition. More details are presented in the Appendix \ref{appendix-compression}.

\begin{table}[htb]
  \begin{center}
    {\small{
\begin{tabular}{ccccc}
\toprule
& IOI & \makecell{IOI \\ + compress.} & \makecell{IOI \\ + random crop} & \makecell{IOI \\ + resize} \\
\midrule
\makecell{RG} & 14.6\% & 12.2\% & 6.30\% & 0.98\% \\
\bottomrule
\end{tabular}
}}
\end{center}
\caption{IOI performance under adversarial purification defences (video compression, random cropping, and resizing). Relative gain averaged for 12 videos.}
\label{tab:disc_compress}
\end{table}



\textbf{Limitations.}
Our method works in a white-box scenario that implies that an attacker knows and has access to the target model. The white-box scenario is less universal than a black-box; however, as described in the introduction, NR quality metrics are usually published as part of the benchmark methodology. We made additional experiments of analysing black-box transferability (Appendix \ref{black-box-transf}) and IOI performance in black-box settings (Appendix \ref{black-box-sett}). We found out low transferability across different models and significant difference in operation speed of white-box and black-box attacks. Black-box attacks are unlikely to be injected into video processing algorithms that compete in quality and speed, which is the scenario we target in our work.
We considered only NR metrics, as FR metrics are much more difficult to attack in real-life scenarios. The robustness of FR metrics has been studied in \cite{ghildyal2023attackPercepMetrics}.

\textbf{Additional experiments}. We made the following additional experiments: experiment to compare the proposed method with prior methods when applying different parameters (Appendix \ref{diff-params}), metric score decreasing experiment (Appendix \ref{score-decr}), IOI attack on segment-level video quality model (Appendix \ref{complex-attack}). We also measured time spending for one-iteration for all methods tested in this paper (Appendix \ref{app:one-it-speed}).

\section{Conclusion}
This paper introduces the IOI adversarial attack on NR image- and video-quality metrics. Its primary objective is to generate imperceptible perturbations for images or videos using only one iteration. Through extensive experiments, we showed that existing methods fail to produce high-fidelity adversarial videos in near real-time scenarios (1 -- 10 fps). In contrast, our proposed method demonstrates better effectiveness at high speed. Subjective and objective comparisons showed that the proposed method produces adversarial images and videos of superior visual quality, achieving the same attack success and speed as prior methods. The proposed attack is a potent tool for experimentally assessing the vulnerability of NR quality metrics. By publishing our method, we provide a tool for verification of NR metrics robustness for benchmark organizers and contribute to the future development of robust image- and video-quality metrics. The proposed method can be used as a part of an adversarial training technique to improve the robustness of image- and video-quality metrics. Our code is openly accessible at \url{https://github.com/katiashh/ioi-attack}.

\section*{Acknowledgements}
The authors would like to thank the CMC Faculty of MSU, especially A.V. Gulyaev for providing the necessary computing resources, which enabled us to undertake some of the calculations for this paper. We also would like to express our gratitude to Mikhail Pautov for discussing the results of this research and assistance in theoretical statements.

The work was supported by a grant for research centers in the field of artificial intelligence, provided by the Analytical Center in accordance with the subsidy agreement (agreement identifier 000000D730321P5Q0002) and the agreement with the Ivannikov Institute for System Programming of dated November 2, 2021 No. 70-2021-00142.

\section*{Impact Statement}
In this paper, we propose a new method that can be used to exploit vulnerabilities in video quality assessment metrics. NR quality assessment is widely used for public competitions. However, little research has been published in this area. Using NR metrics vulnerabilities is profitable for benchmark participants, so they are unlikely to publish their findings. Also, no robust NR metrics have been proposed so far due to the complexity of the task: adversarial training and purification methods reduce the performance of defended methods, significantly reducing the usability of NR metrics as a substitution for subjective tests.

\nocite{langley00}

\bibliography{example_paper}
\bibliographystyle{icml2024}

\newpage
\appendix
\onecolumn

\section{Proof of Theorem 1}
\label{appendix-proof}
\textbf{Theorem 1.} \textit{Let $I \text{ and } I^p$ be original and perturbed image correspondingly, $I^a$ -- IOI adversarial image based on $I^p$ with truncating parameter $f$. Then inequality \ref{eq:formula1} is correct, where $MAE^*(\cdot, \cdot)$ is given by Equation \ref{eq:mae}.}

\begin{equation}
\begin{array}{l}
||I^a - I||_{\infty} \leq (1-f) MAE^*(I^p, I)
\end{array}
\label{eq:formula1}
\end{equation}

\textbf{Proof.} We can estimate the difference between $I^a$ and $I$. Since $||w||_\infty \leq 1$, we can write the following:

\begin{equation}
\begin{array}{l}
I^a = L^{d_f(I)}_c(I) + w H^{d_f(I)}_c(I^p) + (1 - w) H^{d_f(I)}_c(I) \\
I = L^{d_f(I)}_c(I) + H^{d_f(I)}_c(I)) \\
|| I^a - I ||_\infty = ||w (H^{d_f(I)}_c(I^p) - H^{d_f(I)}_c(I))||_\infty \leq  ||w||_\infty  ||(H^{d_f(I)}_c(I^p) - H^{d_f(I)}_c(I))||_\infty \leq \\
\leq ||(H^{d_f(I)}_c(I^p) - H^{d_f(I)}_c(I))||_\infty
\end{array}
\label{eq:124}
\end{equation}

Since FFT and IFFT are linear transformations and indexes $d_f(I)$ for truncating $I^p$ and $I$ are the same: 

\begin{equation}
\begin{array}{l}
H^{d_f(I)}_c(I^p) - H^{d_f(I)}_c(I) = H^{d_f(I)}_c(I^p - I)
\end{array}
\label{eq:122}
\end{equation}

We can write high-frequency component as re-transformed two-dimensional discrete Fourier transform without f\% of highest frequencies and estimate the module of each element of $H^{d_f(I)}_c(I^p - I)$, where $k_r, l_s$ -- indexes of the sorter FFT coefficients, such that $|(I^p - I)^*_{k_n, l_n}| >= |(I^p - I)^*_{k_{n+1}, l_{n+1}}|$ $\forall n$:

\begin{equation}
\begin{array}{l}
|H^{d_f(I)}_c(I^p - I)_{u,v}| 
= \frac{1}{HW}| \displaystyle\sum_{s=f(H-1)(W-1)}^{(H-1)(W-1)} (I^p - I)^*_{k_s, l_s}e^{i2\pi(\frac{k_ru}{H}+\frac{l_sv}{W})}| \leq \\
\leq \frac{1}{HW} \displaystyle\sum_{s=f(H-1)(W-1)}^{(H-1)(W-1)} |(I^p - I)^*_{k_s, l_s}| 
= \frac{1}{HW} \displaystyle\sum_{s=f(H-1)(W-1)}^{(H-1)(W-1)} |(I^p)^*_{k_s, l_s} - (I)^*_{k_s, l_s}| = \beta - \alpha
\end{array}
\end{equation}

where $\alpha$ and $\beta$ are given by Equation \ref{eq:alpha-and-beta2}.

\begin{equation}
\begin{array}{l}
\beta = \frac{1}{HW} \displaystyle\sum_{s=0}^{(H-1)(W-1)} |(I^p)^*_{k_s, l_s} - (I)^*_{k_s, l_s}| \\
\alpha = \frac{1}{HW} \displaystyle\sum_{s=0}^{f(H-1)(W-1)} |(I^p)^*_{k_s, l_s} - (I)^*_{k_s, l_s}|
\end{array}
\label{eq:alpha-and-beta2}
\end{equation}

Considering the facts that $\beta = MAE^*(I^p, I)$ (by definition) and $\alpha >= f MAE^*(I^p, I)$ (since $\alpha$ is the sum of modules of $f$\% highest coefficients and $MAE^*(I^p, I)$ is the sum of all coefficients), we get the resulting estimate:

\begin{equation}
\begin{array}{l}
||I^a - I||_{\infty} \leq MAE^*(I^p, I) - f MAE^*(I^p, I) = (1-f) MAE^*(I^p, I)
\end{array}
\end{equation}

The equation above demonstrates that the proposed method guarantees theoretical restriction of $l_\infty$ norm of the adversarial image, which depends on initial attack strength and $f$ parameter for truncating frequencies. It is worth noting that there was a rough estimate of $||w||_\infty \leq 1$ in Equation \ref{eq:124}. In practice, multiplication on weights highly improves the $l_2$ norm of an adversarial image.

\section{Speed for one iteration}
\label{app:one-it-speed}

Table \ref{tab:times} presents the calculation times of attacks at one iteration that were used for comparison in this paper when targeting the PaQ-2-PiQ \cite{ying2020patches} NR metric on one image from the NIPS2017 dataset \yrcite{nips-dataset} and one video from the DERF dataset \yrcite{derf-dataset}. We measured the calculation time on a server with an NVIDIA Tesla T4 GPU and averaged the results over 20 runs. The AdvJND is notably slower than others due to the computational complexity of calculating JND coefficients.

\begin{table}[htb]
\caption{GPU calculation times of attacks at one iteration when attacking PaQ-2-PiQ \yrcite{ying2020patches} NR metric on images and videos.}
\label{tab:times}
\begin{center}
\begin{small}
\begin{sc}
\begin{tabular}{llr}
\toprule
 Method & \makecell{One iteration \\ time on image} & \makecell{One iteration \\ FPS on video} \\
\midrule
\makecell{FGSM \yrcite{goodfellow_explaining_2015}, SSAH \yrcite{luo_frequency-driven_2022}, \\ Zhang et al. \yrcite{zhang_perceptual_2022}} & 0.025 sec & 8.92 fps \\
NVW \yrcite{karli_improving_2021} & 0.059 sec & 2.78 fps \\
Korhonen et al. \yrcite{korhonen_adversarial_2022} & 0.037 sec & 7.05 fps \\
AdvJND \yrcite{chen_advjnd_2020} & 10.38 sec & 0.01 fps \\
IOI (ours) $_{PyTorch}$ & 0.028 sec & 7.81 fps \\
\bottomrule
\end{tabular}
\end{sc}
\end{small}
\end{center}
\end{table}

\section{Experiment with multiple iterations}
\label{appendix-mult-iters}
We conducted an additional experiment to evaluate the performance of visual-oriented methods employed in this paper in the context of multiple iterations. For $n=10$ and $n=20$ iterations, we run the proposed method with $\epsilon=0.1$ and step size of $\frac{2\epsilon}{n}$. These experiments used the PaQ-2-PiQ NR model \cite{ying2020patches}; the relative gain was 13\% for both 10 and 20 iterations. Utilizing Algorithm \ref{alg:example}, we searched for the minimal $lr$ parameter in other attacks to achieve the same relative gain. Subsequently, we evaluated the visual quality of the resulting adversarial images using four FR metrics: SSIM, PSNR, VIF, and LPIPS. The results for $n=10$ and $n=20$ iterations are presented in Tables \ref{tab:mult-iters10} and \ref{tab:mult-iters20} respectively. At 20 iterations, all methods produce nearly identical results, highlighting the primary strength of the proposed IOI method in its effectiveness in one-iteration settings.

From the results of this experiment, we can conclude that in the setup of multiple iteration attack, there are no significant differences in which method from Table \ref{tab:mult-iters20} to use (except SSAH and AdvJND -- they need more than 20 iterations for convergence). The primary strength of the proposed IOI method is its effectiveness and superiority in one-iteration settings, but in multi-iteration setup, it also shows competitive results.

\begin{table*}[htb]
  \begin{center}
    {\small{
\begin{tabular}{llrrrrr}
\toprule
Method & SSIM $\uparrow$ & PSNR $\uparrow$ & VIF $\uparrow$ & LPIPS $\downarrow$  \\
\midrule
SSAH \yrcite{luo_frequency-driven_2022} & 0.890$\pm$0.005 & 33.6$\pm$0.4 & 0.680$\pm$0.009 & 0.106$\pm$0.005 \\
Zhang et al. \yrcite{zhang_perceptual_2022} & 0.938$\pm$0.003 & 35.7$\pm$0.2 & 0.700$\pm$0.007 & 0.073$\pm$0.004 \\
NVW \yrcite{karli_improving_2021} & 0.957$\pm$0.001 & \underline{37.1$\pm$0.4} & 0.725$\pm$0.006 & 0.055$\pm$0.002 \\
Korhonen et al. \yrcite{korhonen_adversarial_2022} & \textbf{0.974$\pm$0.001} & \textbf{37.5$\pm$0.2} & \underline{0.757$\pm$0.005} & \textbf{0.035$\pm$0.002} \\
AdvJND \yrcite{chen_advjnd_2020} & 0.889$\pm$0.003 & 34.2$\pm$0.2 & 0.546$\pm$0.007 & 0.097$\pm$0.004 \\
IOI (ours) & \underline{0.965$\pm$0.001} & 34.6$\pm$0.2 & \textbf{0.758$\pm$0.004} & \underline{0.043$\pm$0.002} \\
\bottomrule
\end{tabular}
}}
\end{center}
\caption{Comparison results for 10 iterations with relative gain aligning.}
\label{tab:mult-iters10}
\end{table*}

\begin{table*}[htb]
  \begin{center}
    {\small{
\begin{tabular}{llrrrrr}
\toprule
Method & SSIM $\uparrow$ & PSNR $\uparrow$ & VIF $\uparrow$ & LPIPS $\downarrow$  \\
\midrule
SSAH \yrcite{luo_frequency-driven_2022} & 0.949$\pm$0.002 & 36.6$\pm$0.2 & \textbf{0.790$\pm$0.005} & 0.048$\pm$0.002 \\
Zhang et al. \yrcite{zhang_perceptual_2022} & 0.971$\pm$0.001 & 38.1$\pm$0.2 & \underline{0.786$\pm$0.005} & \underline{0.034$\pm$0.002} \\
NVW \yrcite{karli_improving_2021} & 0.970$\pm$0.001 & \textbf{38.7$\pm$0.4} & 0.778$\pm$0.005 & 0.038$\pm$0.002 \\
Korhonen et al. \yrcite{korhonen_adversarial_2022} & \textbf{0.978$\pm$0.001} & \underline{38.2$\pm$0.2} & 0.781$\pm$0.004 & \textbf{0.029$\pm$0.001} \\
AdvJND \yrcite{chen_advjnd_2020} & 0.916$\pm$0.003 & 35.9$\pm$0.2 & 0.613$\pm$0.007 & 0.076$\pm$0.004 \\
IOI (ours) & \underline{0.972$\pm$0.001} & 35.5$\pm$0.2 & 0.779$\pm$0.004 & 0.035$\pm$0.002 \\
\bottomrule
\end{tabular}
}}
\end{center}
\caption{Comparison results for 20 iterations with relative gain aligning.}
\label{tab:mult-iters20}
\end{table*}

\section{Different parameter's comparison}
\label{diff-params}
We made an additional experiment to compare the proposed method with prior methods when applying different parameters. This allows us to compare methods in slightly different attack strengths. Results in the Table \ref{tab:res_dif_params} showed that the proposed method generates images with better visual quality when achieving the same increase in metric score for three different increase levels.

\begin{table*}[!h]
\caption{Experiment of methods comparison under different relative gains, corresponding to different $\epsilon$ levels in the proposed IOI method.}
\label{tab:res_dif_params}
\begin{center}
\begin{small}
\begin{tabular}{llrrrrr}
\toprule
$\epsilon$ & Method & SSIM $\uparrow$ & PSNR $\uparrow$ & VIF $\uparrow$ & LPIPS $\downarrow$  \\
\midrule
\multirow{7}{*}{0.08} & \makecell[l]{FGSM \yrcite{goodfellow_explaining_2015}, SSAH \yrcite{luo_frequency-driven_2022}, \\ Zhang et al. \yrcite{zhang_perceptual_2022}} & 0.934$\pm$0.006 & \underline{36.4$\pm$0.2} & 0.733$\pm$0.009 & 0.082$\pm$0.008 \\
& NVW \yrcite{karli_improving_2021} & \underline{0.940$\pm$0.006} & \textbf{37.4$\pm$0.4} & \underline{0.745$\pm$0.009} & \underline{0.072$\pm$0.007}  \\
& Korhonen et al. \yrcite{korhonen_adversarial_2022} & 0.932$\pm$0.005 & 36.2$\pm$0.2 & 0.727$\pm$0.009 & 0.083$\pm$0.007 \\
& AdvJND \yrcite{chen_advjnd_2020} & 0.812$\pm$0.005 & 31.9$\pm$0.1 & 0.466$\pm$0.007 & 0.151$\pm$0.006  \\
& UAP \yrcite{shumitskaya2022universal} & 0.792$\pm$0.004 & 27.9$\pm$0.2 & 0.432$\pm$0.005 & 0.251$\pm$0.004 \\
& FACPA \yrcite{DBLP:conf/iclr/ShumitskayaAV23} & 0.888$\pm$0.003 & 31.7$\pm$0.2 & 0.586$\pm$0.005 & 0.151$\pm$0.003 \\
& IOI (ours) & \textbf{0.966$\pm$0.002} & 35.5$\pm$0.1 & \textbf{0.786$\pm$0.004} & \textbf{0.037$\pm$0.002} \\
\midrule
\multirow{7}{*}{0.1}& \makecell[l]{FGSM \yrcite{goodfellow_explaining_2015}, SSAH \yrcite{luo_frequency-driven_2022}, \\ Zhang et al. \yrcite{zhang_perceptual_2022}} & 0.884$\pm$0.007 & \underline{33.6$\pm$0.3} & 0.635$\pm$0.010 & 0.134$\pm$0.009  \\
& NVW \yrcite{karli_improving_2021} & \underline{0.897$\pm$0.007} & \textbf{34.7$\pm$0.5} & \underline{0.648$\pm$0.011} & \underline{0.120$\pm$0.008}  \\
& Korhonen et al. \yrcite{korhonen_adversarial_2022} & 0.872$\pm$0.008 & 33.1$\pm$0.3 & 0.617$\pm$0.011 & 0.151$\pm$0.011 \\
& AdvJND \yrcite{chen_advjnd_2020} & 0.740$\pm$0.008 & 29.5$\pm$0.2 & 0.384$\pm$0.008 & 0.208$\pm$0.007 \\
& UAP \yrcite{shumitskaya2022universal} & 0.737$\pm$0.004 & 26.3$\pm$0.2 & 0.371$\pm$0.004 & 0.314$\pm$0.005 \\
& FACPA \yrcite{DBLP:conf/iclr/ShumitskayaAV23} & 0.863$\pm$0.003 & 30.5$\pm$0.2 & 0.539$\pm$0.005 & 0.182$\pm$0.004 \\
& IOI (ours) & \textbf{0.950$\pm$0.002} & 33.4$\pm$0.2 & \textbf{0.695$\pm$0.005} & \textbf{0.059$\pm$0.003} \\
\midrule
\multirow{7}{*}{0.12} & \makecell[l]{FGSM \yrcite{goodfellow_explaining_2015}, SSAH \yrcite{luo_frequency-driven_2022}, \\ Zhang et al. \yrcite{zhang_perceptual_2022}} & 0.789$\pm$0.016 & 30.7$\pm$0.5 & 0.548$\pm$0.015 & 0.274$\pm$0.022 \\
& NVW \yrcite{karli_improving_2021} & \underline{0.795$\pm$0.016} & \underline{31.5$\pm$0.6} & \underline{0.555$\pm$0.016} & 0.264$\pm$0.022  \\
& Korhonen et al. \yrcite{korhonen_adversarial_2022} & 0.774$\pm$0.016 & 30.0$\pm$0.5 & 0.524$\pm$0.016 & 0.295$\pm$0.022 \\
& AdvJND \yrcite{chen_advjnd_2020} & 0.696$\pm$0.007 & 28.4$\pm$0.1 & 0.340$\pm$0.006 & \underline{0.239$\pm$0.008}  \\
& UAP \yrcite{shumitskaya2022universal} & 0.705$\pm$0.004 & 25.4$\pm$0.1 & 0.342$\pm$0.004 & 0.348$\pm$0.007 \\
& FACPA \yrcite{DBLP:conf/iclr/ShumitskayaAV23} & 0.758$\pm$0.003 & 26.9$\pm$0.2 & 0.392$\pm$0.004 & 0.291$\pm$0.005 \\
& IOI (ours) & \textbf{0.936$\pm$0.003} & \textbf{32.2$\pm$0.2} & \textbf{0.681$\pm$0.005} & \textbf{0.077$\pm$0.004} \\
\bottomrule
\end{tabular}
\end{small}
\end{center}
\end{table*}

\section{IOI performance under defences}
\label{appendix-compression}
We conducted additional experiments to evaluate the robustness of the proposed IOI attack to three defences: compression \cite{shaham2018defending}, random crop and resize \cite{shumitskaya2023towards}. Random crop and resize defences were previously studied in \cite{shumitskaya2023towards} for evaluation of UAP \cite{shumitskaya2022universal} against NR quality metrics. Authors showed that random crop and resize to 80\% of the original image/video size help to improve NR quality metric robustness to adversarial attacks without significant loss in correlations with subjective scores. Because of that, we chose parameter 80\% for our experiments. Defended transformation in the case of random crop defence was selecting a random crop of the frame with 80\% of the original frame size. An image was resized to 80\% of the original frame size for resize-based defence. Shaham et al. showed that compression can be used as defence \cite{shaham2018defending}. In our experiment, defended transformation for compression defence was compression with a CRF value 16 using x264 video codec (preset ``Medium'').

For the experiment, we used 12 adversarial videos generated against PaQ-2-PiQ (source videos from the DERF dataset \cite{derf-dataset}) and original ones. We measured relative gain for the IOI attack with and without defences. For relative gain measurement under defence, we calculated PaQ-2-PiQ scores for original and adversarial videos after defended transformation and then, based on these scores, calculated relative gain. Results are presented in the Table \ref{tab:app_compress}. Compression defence reduced relative gain only by 2.4\%. Random crop confuse attack and reduce relative gain approximately two times. Resize defence almost completely mitigate the relative gain, however gain in 1\% still can be significant in benchmarks. From the results of these experiments, we can conclude that the proposed IOI attack is robust to compression and random crop defences but vulnerable to the resize defence.

\begin{table*}[!h]
  \begin{center}
    {\small{
\begin{tabular}{lrrrr}
\toprule
Video & \multicolumn{4}{c}{Relative gain} \\
& IOI & IOI + compression & IOI + random crop & IOI + resize \\
\midrule
Blue Sky & 13.1\% & 11.2\% & 6.17\% & 0.88\% \\
Rush Hour & 23.9\% & 18.0\% & 11.3\% & 0.77\% \\
Old Town Cross & 15.6\% & 13.9\% & 5.33\% & 1.29\% \\
Crowd Run & 15.2\% & 12.4\% & 4.10\% & 0.64\% \\
Aspen & 9.50\% & 6.70\% & 4.11\% & 0.51\% \\
Sunflower & 18.9\% & 15.8\% & 13.3\% & 1.33\% \\
Life & 10.0\% & 7.64\% & 1.76\% & -0.09\% \\
Controlled Burn & 16.2\% & 15.4\% & 7.12\% & 1.85\% \\
Red Kayak & 16.3\% & 14.3\% & 7.08\% & 2.40\% \\
Ducks Take Off & 7.94\% & 6.24\% & 2.80\% & 0.30\% \\
Tractor & 11.6\% & 9.57\% & 6.00\% & 0.74\% \\
Park Joy & 16.8\% & 14.8\% & 6.47\% & 1.10\% \\
\midrule
Mean & 14.6\% & 12.2\% & 6.30\% & 0.98\% \\
\bottomrule
\end{tabular}
}}
\end{center}
\caption{Results of performance proposed IOI attack under compression, random crop and resize defences.}
\label{tab:app_compress}
\end{table*}

\section{Black-box transferability}
\label{black-box-transf}
We conducted an additional experiment to analyse the applicability of IOI in transferable black-box setting. The experiment was organised as follows: for all generated adversarial images from NIPS2017 dataset, we measured PaQ-2-PiQ, Hyper-IQA and TReS, i.e. for adversarial images created to attack PaQ-2-PiQ we also measured quality scores by Hyper-IQA and TReS. Based on these metrics scores, we calculated relative gains. Results are presented in the Table \ref{tab:transferability}. All eight methods tested in the paper showed low transferability to unseen models. Low transferability can be explained by these metrics having completely different architectures: PaQ-2-PiQ employs RoIPool layers, HyperIQA utilizes ResNet50 and TReS is based on transformer architecture. Also, it's important to note that improving transferability for image quality models is more challenging than for classifiers or detectors. Transferability can occur in classification and detection tasks because different classifiers "look" at the same regions of images where classified objects are located \cite{meng2019class}. In contrast, different image quality metrics can look at different regions to estimate the score, which inherently complicates the achievement of transferability. Thus, we can infer that developing imperceptible and, at the same time, transferable one-iteration attacks on video-quality models is a challenging problem that we will consider for further research.

\begin{table*}[h]
\caption{Results of experiment on transferability of methods used in the paper in one-iteration setting.}
\label{tab:transferability}
\label{tab:res}
\begin{center}
\begin{small}
\begin{tabular}{l|rrr|rrr|rrr}
\toprule
Attack & \multicolumn{3}{c|}{\textbf{PaQ-2-PiQ}} & \multicolumn{3}{c|}{\textbf{Hyper-IQA}} & \multicolumn{3}{c}{\textbf{TReS}} \\
Test & \textbf{PaQ-2-PiQ} & Hyper-IQA & TReS & PaQ-2-PiQ & \textbf{Hyper-IQA} & TReS & PaQ-2-PiQ & Hyper-IQA & \textbf{TReS} \\
\midrule
\makecell[l]{FGSM, \\ SSAH, \\ Zhang} & \textbf{7.40\%} & -8.76\% & -14.93\% & 0.16\% & \textbf{1.03\%} & -18.99\% & 0.01\% & -0.39\% & \textbf{4.07\%} \\
NVW & \textbf{7.17\%} & -8.27\% & -14.19\% & 0.13\% & \textbf{0.70\%} & -15.32\% & 0.08\% & 0.09\% & \textbf{4.50\%} \\
Korhonen & \textbf{7.36\%} & -8.85\% & -14.81\% & 0.23\% & \textbf{0.42\%} & -15.97\% & 0.07\% & 0.29\% & \textbf{5.04\%}  \\
AdvJND & \textbf{6.61\%} & -0.57\% & -19.84\% & -0.18\% & \textbf{4.09\%} & -3.96\% & 0.04\% & -0.43\% & \textbf{7.02\%} \\
UAP & \textbf{11.62\%} & -6.03\% & -15.55\% & 3.13\% & \textbf{4.83\%} & 6.65\% & 2.44\% & -0.52\% & \textbf{0.85\%} \\
FACPA & \textbf{9.68\%} & -5.01\% & -9.14\% & -0.30\% & \textbf{1.48\%} & 9.52\% & -6.02\% & -0.51\% & \textbf{1.14\%} \\
IOI (ours) & \textbf{7.33\%} & -3.56\% & -6.25\% & 0.26\% & \textbf{7.86\%} & -1.73\% & 0.39\% & 2.34\% & \textbf{7.40\%} \\
\bottomrule
\end{tabular}
\end{small}
\end{center}
\end{table*}

\section{IOI performance in black-box setting}
\label{black-box-sett}
We conducted an additional experiment aimed to show that the proposed method can be adapted for use in a black-box setting. To do this, we replaced an FGSM-generated perturbation with a black-box-generated perturbation, e.g. Square Attack \cite{andriushchenko2020square}. We will call this modification BB-IOI. It's important to note that such modification transforms the method into a multi-iteration. To verify the efficiency of BB-IOI, we conducted an additional experiment involving the implementation of the BB-IOI attack that consists of two stages: 1) generating the adversarial perturbation in a black-box manner using the Square Attack \cite{andriushchenko2020square} method and 2) processing this perturbation using frequency and weighting modules. As a result, BB-IOI provides high-quality adversarial images by objective metrics. Although the property of imperceptibility remains, it's important to note that the computation complexity of the method has increased, which is common for black-box methods. For attacking PaQ-2-PiQ on the NIPS2017 dataset, BB-IOI achieved an average 1.21\% gain operating at 40 seconds per image. Results are presented in the Table \ref{tab:bb-ioi}. This performance is 6.12\% lower and 1400 times slower than IOI. Furthermore, the proposed IOI method can be adapted for use in transfer-based settings by combining it with transfer-based perturbation generation techniques \cite{long2022frequency}, \cite{li2023frequency}.

\begin{table*}[h]
\caption{Comparison of the proposed method used in white-box setting (IOI) and black-box setting (BB-IOI).}
\label{tab:bb-ioi}
\label{tab:res}
\begin{center}
\begin{small}
\begin{tabular}{lrrrrrr}
\toprule
Method & SSIM $\uparrow$ & PSNR $\uparrow$ & VIF $\uparrow$ & LPIPS $\downarrow$ & Relative gain $\uparrow$ & Time on one image $\downarrow$ \\
\midrule
IOI & 0.945$\pm$0.002 & 33.4$\pm$0.2 & 0.756$\pm$0.005 & 0.059$\pm$0.003  & 7.33\% & 0.028 sec \\
BB-IOI & 0.988$\pm$0.001 & 36.9$\pm$0.1 & 0.815$\pm$0.003 & 0.024$\pm$0.001 & 1.21\% & 40 sec \\
\bottomrule
\end{tabular}
\end{small}
\end{center}
\end{table*}

\section{Metric score decreasing experiment}
\label{score-decr}
In this section, we show the possibility of metrics score decreasing. Given that the proposed method consists of two parts (generating a perturbed image using FGSM and subsequent processing in frequency and weighting modules), changing the optimization direction in FGSM leads to guiding an attack in the opposite direction. We conducted additional experiments to show that the proposed method can decrease PaQ-2-PiQ metric scores. We applied the proposed method to attack it on the same NIPS2017 dataset we used to increase this metric. The results revealed that increasing metric scores yielded a 7.32\% score increase, while decreasing metric scores resulted in a 7.61\% score decrease (almost the same). However, it's worth noting that we focused on increasing metrics’ scores because decreasing quality metrics’ scores holds less practical significance. An attacker can decrease the metrics for quality camouflage \yrcite{Pixel_Privacy}, and it is the only real-life scenario known to the authors.

\section{IOI attack on complex VQA metric}
\label{complex-attack}
The proposed method applies to segment-level video quality models. We conducted an additional experiment targeting the segment-level VQA metric MDTVSFA \cite{li2021unified} to show this. To apply the proposed attack, we first slightly modified MDTVSFA to get access to its gradient (we removed $torch.no\_grad()$ context-manager from the feature extraction module and modified the forward process in the inference model to process batches rather than dictionaries). Then, we applied IOI to attack three videos on a per-frame basis, processing one frame at a time. Subsequently, we constructed adversarial videos from these frames and calculated quality scores using the original segment-level MDTVSFA on these videos. Results are presented in the Table \ref{tab:mdtvsfa}. Remarkably, this approach yielded a significant relative gain, with a 15\% increase in scores. This experiment showed that attacking only spatial features of a VQA metric without accounting for temporal and other features is enough to achieve high attack success.

\begin{table*}[h]
\caption{Experiments of IOI attack targeting segment-level video-quality metric MDTVSFA. The attack was performed per-frame. Resulting gain was calculated using original segment-level MDTVSFA.}
\label{tab:mdtvsfa}
\begin{center}
\begin{small}
\begin{tabular}{lrr}
\toprule
Video & MDTVSFA clean & MDTVSFA IOI attacked \\
\midrule
Blue Sky & 0.544 & 0.659 ($\uparrow$11.5\%) \\
Crowd Run & 0.555 & 0.759 ($\uparrow$20.4\%) \\
Pedestrian Area & 0.584 & 0.737 ($\uparrow$15.3\%) \\
\bottomrule
\end{tabular}
\end{small}
\end{center}
\end{table*}

\section{Video sequences}
\label{sec:appendix-video-desc}
We used the following 12 videos with a resolution of 1280$\times$720 from the DERF dataset \yrcite{derf-dataset}: 
\begin{enumerate}
    \item ``Blue Sky'': \href{https://media.xiph.org/video/derf/y4m/blue_sky_1080p25.y4m}{https://media.xiph.org/video/derf/y4m/blue\_sky\_1080p25.y4m}
    \item ``Aspen'': \href{https://media.xiph.org/video/derf/y4m/aspen_1080p.y4m}{https://media.xiph.org/video/derf/y4m/aspen\_1080p.y4m}
    \item ``Sunflower'': \href{https://media.xiph.org/video/derf/y4m/sunflower_1080p25.y4m}{https://media.xiph.org/video/derf/y4m/sunflower\_1080p25.y4m}
    \item ``Crowd Run'': \href{https://media.xiph.org/video/derf/y4m/crowd_run_1080p50.y4m}{https://media.xiph.org/video/derf/y4m/crowd\_run\_1080p50.y4m}
    \item ``Old Town Cross'': \href{https://media.xiph.org/video/derf/y4m/old_town_cross_1080p50.y4m}{https://media.xiph.org/video/derf/y4m/old\_town\_cross\_1080p50.y4m}
    \item ``Life'': \href{https://media.xiph.org/video/derf/y4m/life_1080p30.y4m}{https://media.xiph.org/video/derf/y4m/life\_1080p30.y4m}
    \item ``Controlled Burn'': \href{https://media.xiph.org/video/derf/y4m/controlled_burn_1080p.y4m}{https://media.xiph.org/video/derf/y4m/controlled\_burn\_1080p.y4m}
    \item ``Rush Hour'': \href{https://media.xiph.org/video/derf/y4m/rush_hour_1080p25.y4m}{https://media.xiph.org/video/derf/y4m/rush\_hour\_1080p25.y4m}
    \item ``Red Kayak'': \href{https://media.xiph.org/video/derf/y4m/red_kayak_1080p.y4m}{https://media.xiph.org/video/derf/y4m/red\_kayak\_1080p.y4m}
    \item ``Ducks Take Off'': \href{https://media.xiph.org/video/derf/y4m/ducks_take_off_1080p50.y4m}{https://media.xiph.org/video/derf/y4m/ducks\_take\_off\_1080p50.y4m}
    \item ``Tractor'': \href{https://media.xiph.org/video/derf/y4m/tractor_1080p25.y4m}{https://media.xiph.org/video/derf/y4m/tractor\_1080p25.y4m}
    \item ``Park Joy'': \href{https://media.xiph.org/video/derf/y4m/park_joy_1080p50.y4m}{https://media.xiph.org/video/derf/y4m/park\_joy\_1080p50.y4m}
\end{enumerate} 

\begin{figure*}[h!]
 \begin{center}
 \centerline{\includegraphics[width=2.5in]{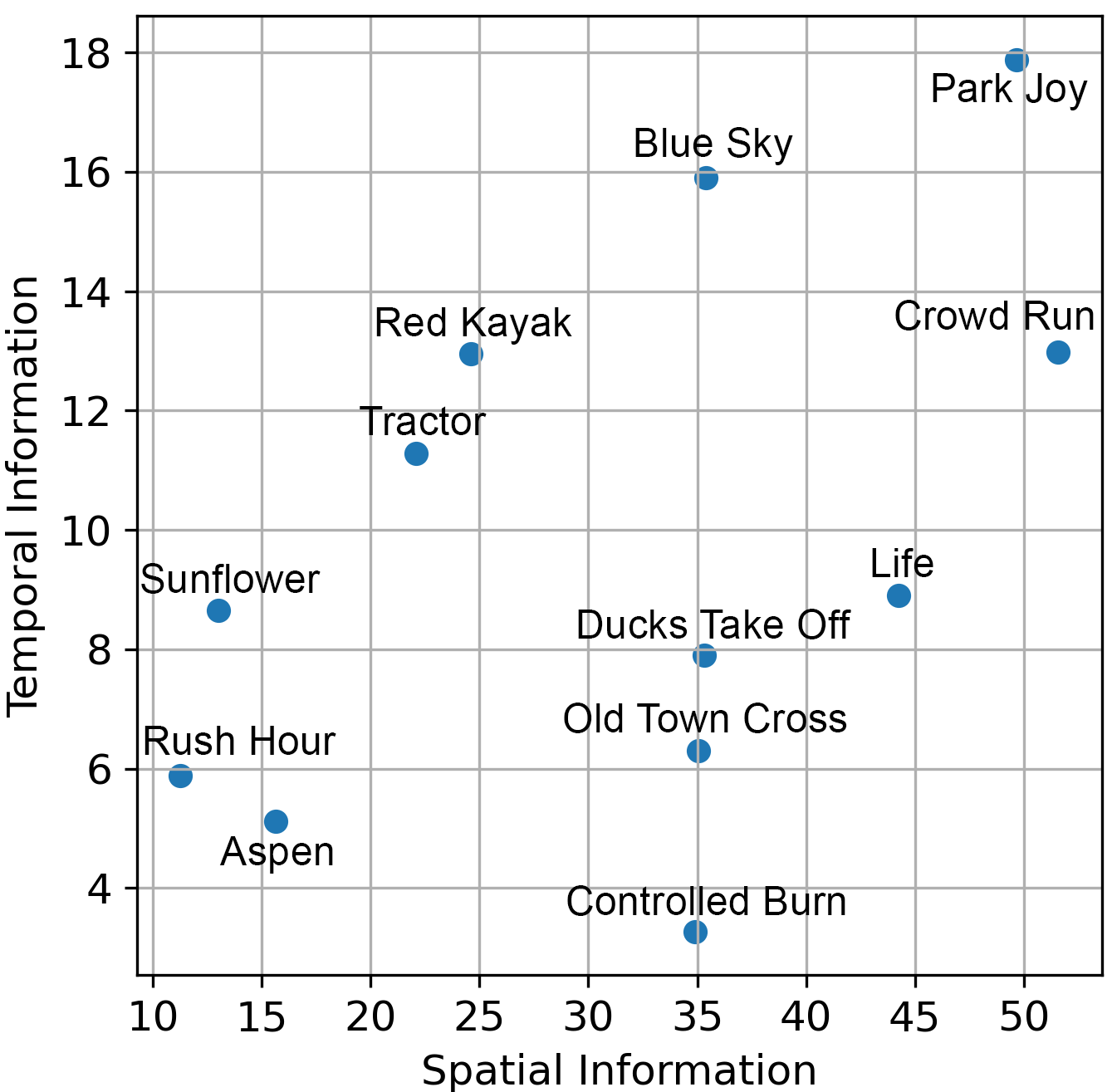}}
 \caption{Spatial and temporal information for videos.}
 \label{fig:si_ti}
 \end{center}
 \end{figure*}

\begin{figure*}[h!]
 \begin{center}
 \centerline{\includegraphics[width=\linewidth]{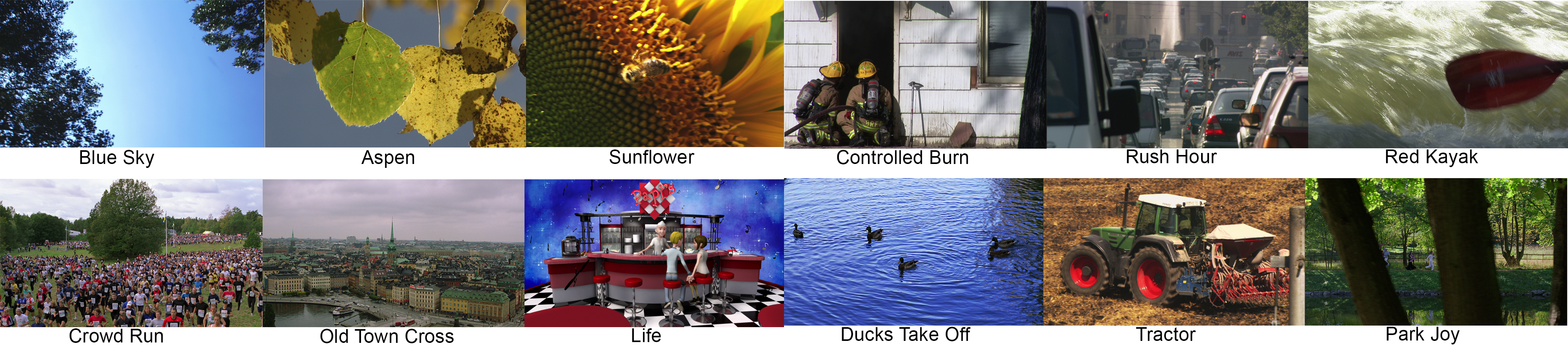}}
 \caption{First frames of videos.}
 \label{fig:previews}
 \end{center}
 \end{figure*}

We extracted 75 frames from each original video and saved an attacked video with a frame rate of 25 frames per second, resulting in videos with a duration of 3 seconds. Figure \ref{fig:si_ti} contains spatial and temporal information for these videos. Figure \ref{fig:previews} contains the first frames of videos. 

\section{Subjective experiment setup}
\label{appendix-subj-comp}

To derive subjective scores for adversarial videos, we conducted a crowd-sourced subjective comparison on the Subjectify.us service \cite{subjectify}. 

For comparison, we compressed all videos, including the original ones, using the x264 video codec with a CRF value of 16 (preset ``Medium''). Each pair shown to participants consisted of two samples of the same source video attacked by various attack methods. Each participant was presented with a random pair of videos sequentially and was asked to choose the video with the superior visual quality. An option ``Can't choose'' was also provided. Videos were pre-loaded in the browser to prevent delays in playback, and participants had the flexibility to replay the videos multiple times. Each participant compared 12 video pairs, of which two were for verification. Answers from 200 participants who failed the verification were excluded. 

We collected 8220 responses from 685 successful participants and calculated subjective scores using the Bradley-Terry model \cite{bradley1952rank}. The average payment to crowdworkers per a pair of sequences was \$0.05. We estimate the overall cost of the subjective tests was \$410. Figure \ref{fig:subj_main_sch} presents the subjective experiment’s general process.

\begin{figure*}[h!]
\begin{center}
\centerline{\includegraphics[width=3.0in]{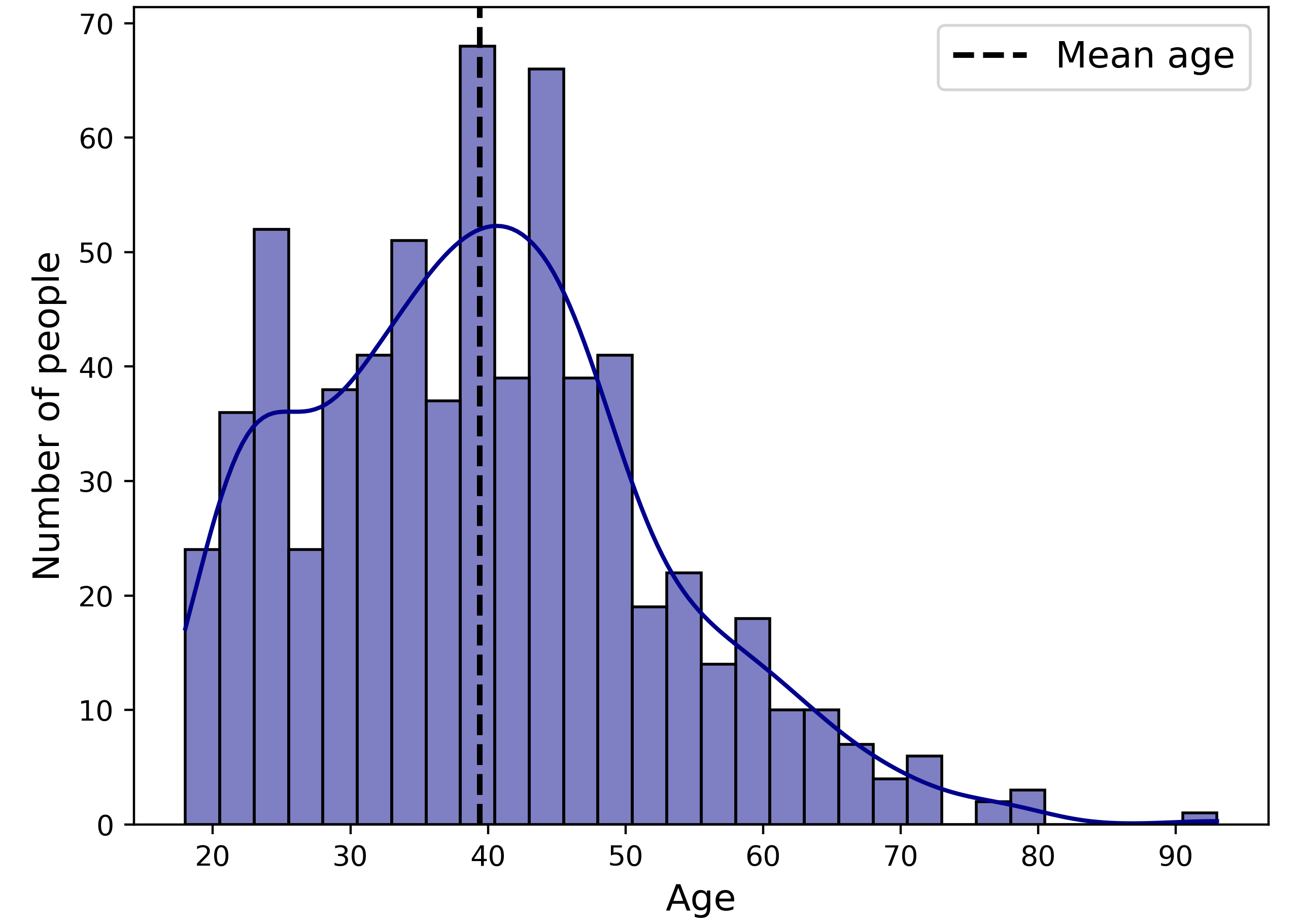}}
\caption{Age distribution of crowdworkers.}
\label{fig:age-distr}
\end{center}
\end{figure*}

Details about the crowdsourced study:
\begin{enumerate}
    \item Screen resolutions were from 360$\times$800 to 3440$\times$1440. Table \ref{tab:subj-ages} shows the most popular.
    \item Participants were from 31 countries.
    \item Participant ages ranged from 18 to 93, with an average of 39. Figure \ref{fig:age-distr} shows the distribution.
\end{enumerate}

\textbf{Command line for encoding.} Given the directory of video frames in PNG format (set of images 000.png, 001.png,
..., 074.png) we run the following FFmpeg command line:
\begin{verbatim}
ffmpeg -pattern_type glob -i *.png -c:v libx264 -crf 16 -pix_fmt yuv420p res.mp4
\end{verbatim}

\begin{figure*}[htb]
\begin{center}
\centerline{\includegraphics[width=\linewidth]{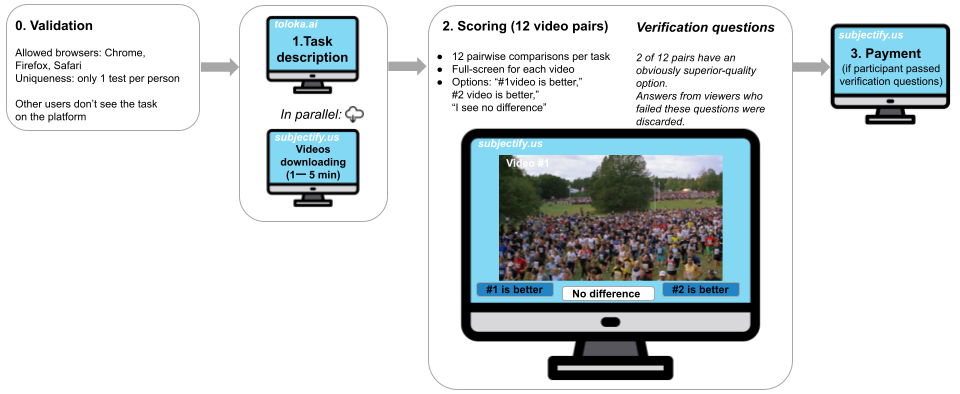}}
\caption{Subjective-assessment scheme.}
\label{fig:subj_main_sch}
\end{center}
\end{figure*}

\section{Per-video results}
\label{sec:appendix-per-video}

Tables \ref{tab:res_vid1}, \ref{tab:res_vid2}, \ref{tab:res_vid3}, \ref{tab:res_vid4}, \ref{tab:res_vid5}, \ref{tab:res_vid6}, \ref{tab:res_vid7}, \ref{tab:res_vid8}, \ref{tab:res_vid9}, \ref{tab:res_vid10}, \ref{tab:res_vid11}, \ref{tab:res_vid12} contain results of subjective comparison of proposed IOI adversarial attack with eight prior attacks for each video.

\section{Attack examples on images and videos}

IOI adversarial images and videos are available in the zip archive: \url{https://drive.google.com/file/d/1nrvV70Q4W0vh-2FdWrXHUhMDzYcI6zY1/view?usp=sharing}.

\begin{table*}[htb]
  \begin{center}
    {\small{
\begin{tabular}{lr}
\toprule
Resolution & Number of users  \\
\midrule
1920$\times$1080 & 194 \\
1366$\times$768 & 167 \\
1536$\times$864 & 100 \\
1280$\times$1024 & 39 \\
1600$\times$900 & 35 \\
1280$\times$720 & 22 \\
1440$\times$900 & 19 \\
1024$\times$768 & 11 \\
2560$\times$1440 & 10 \\
1680$\times$1050 & 10 \\
1360$\times$768 & 10 \\
\bottomrule
\end{tabular}
}}
\end{center}
\caption{Most popular screen resolutions among crowdworkers.}
\label{tab:subj-ages}
\end{table*}

\begin{table*}[h!]
  \begin{center}
    {\small{
\begin{tabular}{llrrrrr}
\toprule
Method & SSIM $\uparrow$ & PSNR $\uparrow$ & VIF $\uparrow$ & LPIPS $\downarrow$ & \makecell{Subjective \\ score $\uparrow$} \\
\midrule
FGSM \yrcite{goodfellow_explaining_2015}, SSAH \yrcite{luo_frequency-driven_2022}, Zhang et al. \yrcite{zhang_perceptual_2022} & 0.739 & 30.8 & 0.390 & 0.352 & 1.44$\pm$0.66 \\
NVW \yrcite{karli_improving_2021} & 0.719 & 30.1 & 0.370 & 0.364 & 1.91$\pm$0.64 \\
Korhonen et al. \yrcite{korhonen_adversarial_2022} & 0.699 & 29.8 & 0.361 & 0.380 & 1.56$\pm$0.65 \\
AdvJND \yrcite{chen_advjnd_2020} & 0.795 & \underline{33.9} & 0.422 & \underline{0.247} & \underline{2.20$\pm$0.63} \\
UAP \yrcite{shumitskaya2022universal} & 0.809 & 31.6 & 0.453 & 0.338 & 0.70$\pm$0.70 \\
FACPA \yrcite{DBLP:conf/iclr/ShumitskayaAV23} & \underline{0.890} & \textbf{34.4} & \underline{0.557} & 0.253 & 1.11$\pm$0.67 \\
IOI (ours) & \textbf{0.956} & \underline{33.9} & \textbf{0.649} & \textbf{0.048} & \textbf{4.05$\pm$0.62} \\
\bottomrule
\end{tabular}
}}
\end{center}
\caption{Comparison results on the ``Blue Sky'' video and PaQ-2-PiQ attacked model \cite{ying2020patches} with relative gain aligning. FR quality metric score for video is calculated as a mean of quality scores on each frame.}
\label{tab:res_vid1}
\end{table*}

\begin{table*}[h!]
  \begin{center}
    {\small{
\begin{tabular}{llrrrrr}
\toprule
Method & SSIM $\uparrow$ & PSNR $\uparrow$ & VIF $\uparrow$ & LPIPS $\downarrow$ & \makecell{Subjective \\ score $\uparrow$} \\
\midrule
FGSM \yrcite{goodfellow_explaining_2015}, SSAH \yrcite{luo_frequency-driven_2022}, Zhang et al. \yrcite{zhang_perceptual_2022} & 0.956 &  40.0 & 0.720 & 0.064 & 2.88$\pm$0.47 \\
NVW \yrcite{karli_improving_2021} & \underline{0.959} & \underline{40.3} & \underline{0.733} & 0.057 & \underline{3.13$\pm$0.48} \\
Korhonen et al. \yrcite{korhonen_adversarial_2022} & 0.957 & 40.2 & 0.725 & 0.062 & 2.91$\pm$0.48 \\
AdvJND \yrcite{chen_advjnd_2020} & 0.950 & \textbf{41.1} & 0.687 & \underline{0.052} & 2.57$\pm$0.48 \\
UAP \yrcite{shumitskaya2022universal} & 0.842 & 32.8 & 0.465 & 0.264 & 0.65$\pm$0.65 \\
FACPA \yrcite{DBLP:conf/iclr/ShumitskayaAV23} & 0.915 & 36.0 & 0.605 & 0.174 & 1.18$\pm$0.58 \\
IOI (ours) & \textbf{0.981} & 38.3 & \textbf{0.823} & \textbf{0.044} & \textbf{3.33$\pm$0.48} \\
\bottomrule
\end{tabular}
}}
\end{center}
\caption{Comparison results on the ``Aspen'' video and PaQ-2-PiQ attacked model \cite{ying2020patches} with relative gain aligning. FR quality metric score for video is calculated as a mean of quality scores on each frame.}
\label{tab:res_vid2}
\end{table*}

\begin{table*}[h!]
  \begin{center}
    {\small{
\begin{tabular}{llrrrrr}
\toprule
Method & SSIM $\uparrow$ & PSNR $\uparrow$ & VIF $\uparrow$ & LPIPS $\downarrow$ & \makecell{Subjective \\ score $\uparrow$} \\
\midrule
FGSM \yrcite{goodfellow_explaining_2015}, SSAH \yrcite{luo_frequency-driven_2022}, Zhang et al. \yrcite{zhang_perceptual_2022} & 0.919 & \underline{37.2} & 0.604 & 0.206 & 2.71$\pm$0.52 \\
NVW \yrcite{karli_improving_2021} & \underline{0.924} & \textbf{37.4} & \underline{0.619} & \underline{0.195} & \underline{2.96$\pm$0.52} \\
Korhonen et al. \yrcite{korhonen_adversarial_2022} & 0.844 & 34.0 & 0.465 & 0.326 & 2.28$\pm$0.53 \\
AdvJND \yrcite{chen_advjnd_2020} & 0.741 & 32.7 & 0.328 & 0.365 & 1.57$\pm$0.56 \\
UAP \yrcite{shumitskaya2022universal} & 0.846 & 32.8 & 0.463 & 0.345 & 0.65$\pm$0.65 \\
FACPA \yrcite{DBLP:conf/iclr/ShumitskayaAV23} & 0.922 & 35.9 & 0.618 & 0.209 & 1.20$\pm$0.60 \\
IOI (ours) & \textbf{0.946} & 36.9 & \textbf{0.690} & \textbf{0.181} & \textbf{3.39$\pm$0.52} \\
\bottomrule
\end{tabular}
}}
\end{center}
\caption{Comparison results on the ``Sunflower'' video and PaQ-2-PiQ attacked model \cite{ying2020patches} with relative gain aligning. FR quality metric score for video is calculated as a mean of quality scores on each frame.}
\label{tab:res_vid3}
\end{table*}

\begin{table*}[h!]
  \begin{center}
    {\small{
\begin{tabular}{llrrrrr}
\toprule
Method & SSIM $\uparrow$ & PSNR $\uparrow$ & VIF $\uparrow$ & LPIPS $\downarrow$ & \makecell{Subjective \\ score $\uparrow$} \\
\midrule
FGSM \yrcite{goodfellow_explaining_2015}, SSAH \yrcite{luo_frequency-driven_2022}, Zhang et al. \yrcite{zhang_perceptual_2022} & 0.921 & \underline{33.3} & 0.689 & 0.071 & \underline{3.61$\pm$0.51} \\
NVW \yrcite{karli_improving_2021} & \underline{0.924} & \textbf{33.4} & \underline{0.694} & \underline{0.068} & 3.16$\pm$0.51 \\
Korhonen et al. \yrcite{korhonen_adversarial_2022} & 0.922 & \textbf{33.4} & 0.693 & 0.070 & 3.51$\pm$0.51 \\
AdvJND \yrcite{chen_advjnd_2020} & 0.844 & 31.7 & 0.525 & 0.089 & 2.12$\pm$0.56 \\
UAP \yrcite{shumitskaya2022universal} & 0.798 & 27.2 & 0.465 & 0.204 & 0.73$\pm$0.73 \\
FACPA \yrcite{DBLP:conf/iclr/ShumitskayaAV23} & 0.885 & 30.5 & 0.604 & 0.140 & 1.44$\pm$0.64 \\
IOI (ours) & \textbf{0.951} & 32.4 & \textbf{0.695} & \textbf{0.039} & \textbf{4.00$\pm$0.51} \\
\bottomrule
\end{tabular}
}}
\end{center}
\caption{Comparison results on the ``Crowd Run'' video and PaQ-2-PiQ attacked model \cite{ying2020patches} with relative gain aligning. FR quality metric score for video is calculated as a mean of quality scores on each frame.}
\label{tab:res_vid4}
\end{table*}

\begin{table*}[h!]
  \begin{center}
    {\small{
\begin{tabular}{llrrrrr}
\toprule
Method & SSIM $\uparrow$ & PSNR $\uparrow$ & VIF $\uparrow$ & LPIPS $\downarrow$ & \makecell{Subjective \\ score $\uparrow$} \\
\midrule
FGSM \yrcite{goodfellow_explaining_2015}, SSAH \yrcite{luo_frequency-driven_2022}, Zhang et al. \yrcite{zhang_perceptual_2022} & 0.854 & 31.8 & 0.512 & 0.151 & 2.77$\pm$1.16 \\
NVW \yrcite{karli_improving_2021} & 0.856 & 31.9 & 0.517 & 0.147 & 2.92$\pm$1.16 \\
Korhonen et al. \yrcite{korhonen_adversarial_2022} & 0.854 & 31.9 & 0.513 & 0.150 & \underline{3.05$\pm$1.16} \\
AdvJND \yrcite{chen_advjnd_2020} & 0.844 & \underline{33.5} & 0.471 & \underline{0.120} & 1.90$\pm$1.18 \\
UAP \yrcite{shumitskaya2022universal} & 0.827 & 30.4 & 0.473 & 0.237 & 1.21$\pm$1.21 \\
FACPA \yrcite{DBLP:conf/iclr/ShumitskayaAV23} & \textbf{0.913} & \textbf{34.1} & \textbf{0.623} & 0.168 & 2.59$\pm$1.16 \\
IOI (ours) & \underline{0.908} & 31.1 & \underline{0.539} & \textbf{0.118} & \textbf{3.61$\pm$1.15} \\
\bottomrule
\end{tabular}
}}
\end{center}
\caption{Comparison results on the ``Old Town Cross'' video and PaQ-2-PiQ attacked model \cite{ying2020patches} with relative gain aligning. FR quality metric score for video is calculated as a mean of quality scores on each frame.}
\label{tab:res_vid5}
\end{table*}

\begin{table*}[h!]
  \begin{center}
    {\small{
\begin{tabular}{llrrrrr}
\toprule
Method & SSIM $\uparrow$ & PSNR $\uparrow$ & VIF $\uparrow$ & LPIPS $\downarrow$ & \makecell{Subjective \\ score $\uparrow$} \\
\midrule
FGSM \yrcite{goodfellow_explaining_2015}, SSAH \yrcite{luo_frequency-driven_2022}, Zhang et al. \yrcite{zhang_perceptual_2022} & 0.737 & 29.7 & 0.419 & 0.216 & 2.44$\pm$0.66 \\
NVW \yrcite{karli_improving_2021} & 0.818 & 31.4 &  0.497 & 0.153 & 2.83$\pm$0.64 \\
Korhonen et al. \yrcite{korhonen_adversarial_2022} & \underline{0.873} & 33.8 & \underline{0.588} & 0.109 & \underline{4.14$\pm$0.59} \\
AdvJND \yrcite{chen_advjnd_2020} & 0.832 & \underline{34.1} & 0.512 & \underline{0.092} & 3.57$\pm$0.61 \\
UAP \yrcite{shumitskaya2022universal} & 0.717 & 28.1 & 0.396 & 0.264 & 0.79$\pm$0.79 \\
FACPA \yrcite{DBLP:conf/iclr/ShumitskayaAV23} & 0.814 & 30.7 & 0.502 & 0.188 & 1.50$\pm$0.72 \\
IOI (ours) & \textbf{0.936} & \textbf{34.8} & \textbf{0.668} & \textbf{0.063} & \textbf{4.89$\pm$0.59} \\
\bottomrule
\end{tabular}
}}
\end{center}
\caption{Comparison results on the ``Life'' video and PaQ-2-PiQ attacked model \cite{ying2020patches} with relative gain aligning. FR quality metric score for video is calculated as a mean of quality scores on each frame.}
\label{tab:res_vid6}
\end{table*}

\begin{table*}[h!]
  \begin{center}
    {\small{
\begin{tabular}{llrrrrr}
\toprule
Method & SSIM $\uparrow$ & PSNR $\uparrow$ & VIF $\uparrow$ & LPIPS $\downarrow$ & \makecell{Subjective \\ score $\uparrow$} \\
\midrule
FGSM \yrcite{goodfellow_explaining_2015}, SSAH \yrcite{luo_frequency-driven_2022}, Zhang et al. \yrcite{zhang_perceptual_2022} & 0.833 & 30.7 & 0.532 & 0.256 & 1.64$\pm$0.52 \\
NVW \yrcite{karli_improving_2021} & 0.853 & 31.1 & 0.553 & 0.225 & \underline{1.93$\pm$0.52} \\
Korhonen et al. \yrcite{korhonen_adversarial_2022} & 0.777 & 28.8 &  0.458 & 0.328 & 1.18$\pm$0.54 \\
AdvJND \yrcite{chen_advjnd_2020} & 0.833 &  \underline{32.7} & 0.508 & \underline{0.176} & 1.54$\pm$0.52 \\
UAP \yrcite{shumitskaya2022universal} & 0.798 & 28.8 & 0.476 & 0.314 & 0.57$\pm$0.57 \\
FACPA \yrcite{DBLP:conf/iclr/ShumitskayaAV23} & \underline{0.903} & \textbf{32.9} & \underline{0.648} & 0.195 & 1.01$\pm$0.55 \\
IOI (ours) & \textbf{0.932} & 32.5 & \textbf{0.652} & \textbf{0.117} & \textbf{2.90$\pm$0.53} \\
\bottomrule
\end{tabular}
}}
\end{center}
\caption{Comparison results on the ``Controlled Burn'' video and PaQ-2-PiQ attacked model \cite{ying2020patches} with relative gain aligning. FR quality metric score for video is calculated as a mean of quality scores on each frame.}
\label{tab:res_vid7}
\end{table*}

\begin{table*}[h!]
  \begin{center}
    {\small{
\begin{tabular}{llrrrrr}
\toprule
Method & SSIM $\uparrow$ & PSNR $\uparrow$ & VIF $\uparrow$ & LPIPS $\downarrow$ & \makecell{Subjective \\ score $\uparrow$} \\
\midrule
FGSM \yrcite{goodfellow_explaining_2015}, SSAH \yrcite{luo_frequency-driven_2022}, Zhang et al. \yrcite{zhang_perceptual_2022} & 0.956 & 40.0 & 0.685 & 0.093 & 3.04$\pm$0.49 \\
NVW \yrcite{karli_improving_2021} & \underline{0.960} & \underline{40.4} & \underline{0.704} & \underline{0.082} & 3.08$\pm$0.49 \\
Korhonen et al. \yrcite{korhonen_adversarial_2022} & 0.959 & \underline{40.4} & 0.701 & 0.084 & \underline{3.39$\pm$0.49} \\
AdvJND \yrcite{chen_advjnd_2020} & \textbf{0.974} & \textbf{44.1} & \textbf{0.767} & \textbf{0.035} & \textbf{3.51$\pm$0.49} \\
UAP \yrcite{shumitskaya2022universal} & 0.838 & 32.8 & 0.421 & 0.357 & 0.71$\pm$0.71 \\
FACPA \yrcite{DBLP:conf/iclr/ShumitskayaAV23} & 0.913 & 36.0 & 0.557 & 0.247 & 1.58$\pm$0.59 \\
IOI (ours) & 0.952 & 38.5 & 0.695 & 0.118 & 3.14$\pm$0.49 \\
\bottomrule
\end{tabular}
}}
\end{center}
\caption{Comparison results on the ``Rush Hour'' video and PaQ-2-PiQ attacked model \cite{ying2020patches} with relative gain aligning. FR quality metric score for video is calculated as a mean of quality scores on each frame.}
\label{tab:res_vid8}
\end{table*}

\begin{table*}[h!]
  \begin{center}
    {\small{
\begin{tabular}{llrrrrr}
\toprule
Method & SSIM $\uparrow$ & PSNR $\uparrow$ & VIF $\uparrow$ & LPIPS $\downarrow$ & \makecell{Subjective \\ score $\uparrow$} \\
\midrule
FGSM \yrcite{goodfellow_explaining_2015}, SSAH \yrcite{luo_frequency-driven_2022}, Zhang et al. \yrcite{zhang_perceptual_2022} & 0.790 & 30.7 & 0.466 & 0.333 & 1.36$\pm$0.56 \\
NVW \yrcite{karli_improving_2021} & 0.843 & 32.3 & 0.535 & 0.249 & \underline{2.20$\pm$0.54} \\
Korhonen et al. \yrcite{korhonen_adversarial_2022} & 0.792 & 30.7 & 0.469 & 0.329 & 1.84$\pm$0.54 \\
AdvJND \yrcite{chen_advjnd_2020} & 0.759 & 31.7 & 0.393 & 0.272 & 1.42$\pm$0.56 \\
UAP \yrcite{shumitskaya2022universal} & 0.840 & 31.5 & 0.517 & 0.280 & 0.61$\pm$0.61 \\
FACPA \yrcite{DBLP:conf/iclr/ShumitskayaAV23} & \underline{0.900} & \underline{34.0} & \underline{0.638} & \underline{0.219} & 1.20$\pm$0.57 \\
IOI (ours) & \textbf{0.942} & \textbf{34.3} & \textbf{0.675} & \textbf{0.116} & \textbf{3.38$\pm$0.55} \\
\bottomrule
\end{tabular}
}}
\end{center}
\caption{Comparison results on the ``Red Kayak'' video and PaQ-2-PiQ attacked model \cite{ying2020patches} with relative gain aligning. FR quality metric score for video is calculated as a mean of quality scores on each frame.}
\label{tab:res_vid9}
\end{table*}

\begin{table*}[h!]
  \begin{center}
    {\small{
\begin{tabular}{llrrrrr}
\toprule
Method & SSIM $\uparrow$ & PSNR $\uparrow$ & VIF $\uparrow$ & LPIPS $\downarrow$ & \makecell{Subjective \\ score $\uparrow$} \\
\midrule
FGSM \yrcite{goodfellow_explaining_2015}, SSAH \yrcite{luo_frequency-driven_2022}, Zhang et al. \yrcite{zhang_perceptual_2022} & 0.904 & 30.6 & 0.549 & 0.172 & 3.42$\pm$0.56 \\
NVW \yrcite{karli_improving_2021} & \underline{0.906} & 30.7 & 0.551 & 0.170 & \underline{3.62$\pm$0.56} \\
Korhonen et al. \yrcite{korhonen_adversarial_2022} & 0.905 & 30.7 & \underline{0.553} & 0.168 & 3.42$\pm$0.56 \\
AdvJND \yrcite{chen_advjnd_2020} & 0.895 & \underline{32.0} & 0.515 & \underline{0.103} & 3.04$\pm$0.57 \\
UAP \yrcite{shumitskaya2022universal} & 0.826 & 26.0 & 0.400 & 0.346 & 0.77$\pm$0.77 \\
FACPA \yrcite{DBLP:conf/iclr/ShumitskayaAV23} & 0.880 & 28.6 & 0.491 & 0.194 & 1.48$\pm$0.68 \\
IOI (ours) & \textbf{0.963} & \textbf{34.1} & \textbf{0.693} & \textbf{0.055} & \textbf{4.09$\pm$0.55} \\
\bottomrule
\end{tabular}
}}
\end{center}
\caption{Comparison results on the ``Ducks Take Off'' video and PaQ-2-PiQ attacked model \cite{ying2020patches} with relative gain aligning. FR quality metric score for video is calculated as a mean of quality scores on each frame.}
\label{tab:res_vid10}
\end{table*}

\begin{table*}[h!]
  \begin{center}
    {\small{
\begin{tabular}{llrrrrr}
\toprule
Method & SSIM $\uparrow$ & PSNR $\uparrow$ & VIF $\uparrow$ & LPIPS $\downarrow$ & \makecell{Subjective \\ score $\uparrow$} \\
\midrule
FGSM \yrcite{goodfellow_explaining_2015}, SSAH \yrcite{luo_frequency-driven_2022}, Zhang et al. \yrcite{zhang_perceptual_2022} & 0.893 & 33.3 & 0.598 & 0.182 & 3.48$\pm$0.56 \\
NVW \yrcite{karli_improving_2021} & \underline{0.902} & 33.5 & 0.612 & 0.168 & \underline{4.29$\pm$0.54} \\
Korhonen et al. \yrcite{korhonen_adversarial_2022} & 0.899 & 33.7 & \underline{0.615} & 0.168 & 3.56$\pm$0.56 \\
AdvJND \yrcite{chen_advjnd_2020} & 0.892 & \textbf{35.0} & 0.584 & \textbf{0.111} & 3.66$\pm$0.55 \\
UAP \yrcite{shumitskaya2022universal} & 0.800 & 28.8 & 0.444 & 0.333 & 0.86$\pm$0.86 \\
FACPA \yrcite{DBLP:conf/iclr/ShumitskayaAV23} & 0.877 & 31.5 & 0.561 & 0.221 & 1.75$\pm$0.73 \\
IOI (ours) & \textbf{0.944} & \underline{34.7} & \textbf{0.696} & \underline{0.119} & \textbf{4.72$\pm$0.54} \\
\bottomrule
\end{tabular}
}}
\end{center}
\caption{Comparison results on the ``Tractor'' video and PaQ-2-PiQ attacked model \cite{ying2020patches} with relative gain aligning. FR quality metric score for video is calculated as a mean of quality scores on each frame.}
\label{tab:res_vid11}
\end{table*}

\begin{table*}[h!]
  \begin{center}
    {\small{
\begin{tabular}{llrrrrr}
\toprule
Method & SSIM $\uparrow$ & PSNR $\uparrow$ & VIF $\uparrow$ & LPIPS $\downarrow$ & \makecell{Subjective \\ score $\uparrow$} \\
\midrule
FGSM \yrcite{goodfellow_explaining_2015}, SSAH \yrcite{luo_frequency-driven_2022}, Zhang et al. \yrcite{zhang_perceptual_2022} & 0.807 & 29.7 & 0.491 & 0.240 & \underline{2.55$\pm$0.65} \\
NVW \yrcite{karli_improving_2021} & 0.784 & 28.8 & 0.458 & 0.263 & 2.07$\pm$0.66 \\
Korhonen et al. \yrcite{korhonen_adversarial_2022} & 0.780 & 28.8 & 0.456 & 0.269 & 2.23$\pm$0.65 \\
AdvJND \yrcite{chen_advjnd_2020} & 0.812 & \textbf{31.8} & 0.485 & \underline{0.175} & 1.94$\pm$0.66 \\
UAP \yrcite{shumitskaya2022universal} & 0.764 & 27.3 & 0.424 & 0.325 & 0.74$\pm$0.74 \\
FACPA \yrcite{DBLP:conf/iclr/ShumitskayaAV23} & \underline{0.834} & 29.7 & \underline{0.525} & 0.273 & 1.41$\pm$0.69 \\
IOI (ours) & \textbf{0.881} & \underline{30.2} & \textbf{0.549} & \textbf{0.161} & \textbf{3.29$\pm$0.63} \\
\bottomrule
\end{tabular}
}}
\end{center}
\caption{Comparison results on the ``Park Joy'' video and PaQ-2-PiQ attacked model \cite{ying2020patches} with relative gain aligning. FR quality metric score for video is calculated as a mean of quality scores on each frame.}
\label{tab:res_vid12}
\end{table*}

\end{document}